\documentclass[useAMS,usenatbib]{mn2e}
\usepackage{graphicx}
\usepackage[dvips]{color}
\usepackage{txfonts}
\usepackage{multirow}
\usepackage{amsfonts}
\usepackage{amssymb}
\usepackage{natbib}

\newcommand{\te}{$T_{\rm eff}$}
\newcommand{\logg}{$\log{g}$}
\newcommand{\vsini}{$v\sin{i}$}
\newcommand{\vmicro}{$v_{\rm turb}$}
\newcommand{\kms}{km\,s$^{-1}$}
\newcommand{\cd}{d$^{-1}$}
\newcommand{\mhz}{$\mu$Hz}

\title[Stellar modelling of Spica]{Stellar modelling of Spica, a high-mass spectroscopic binary with a $\beta$\,Cep variable primary component\thanks{Based on data from the MOST satellite, a Canadian Space Agency mission, jointly operated by Dynacon Inc., the University of Toronto Institute for Aerospace Studies and the University of British Columbia, with the assistance of the University of Vienna.}}

\author[A. Tkachenko et al.]{A.\ Tkachenko,$^{1,}$\thanks{Postdoctoral Fellow of the Research Foundation Flanders} J.\ M.\ Matthews,$^2$, C.\ Aerts,$^{1,3}$ K.\ Pavlovski,$^4$ P.~\ I.\ P\'{a}pics,$^1$\thanks{Postdoctoral Fellow of the Research Foundation Flanders}\newauthor
        K. Zwintz,$^5$ C.\ Cameron,$^{2,6}$ G.\ A.\ H.\ Walker,$^2$ R.\ Kuschnig,$^7$ P.\ Degroote,$^{1,}$ \newauthor
        J.\ Debosscher,$^1$ E.\ Moravveji,$^{1,}$ V.\ Kolbas,$^4$ D.\ B.\ Guenther,$^8$ A.\ F.\ J.\ Moffat,$^9$ \newauthor
        J.\ F.\ Rowe,$^{10}$ S.\ M.\ Rucinski,$^{11}$ D.\ Sasselov,$^{12}$ and W.\ W.\ Weiss,$^7$\\
  $^1$Instituut voor Sterrenkunde, KU Leuven, Celestijnenlaan 200D, B-3001 Leuven, Belgium\\
  $^2$Department of Physics and Astronomy, University of British Columbia, 6224 Agricultural Road, Vancouver, BC V6T 1Z1, Canada\\
  $^3$Department of Astrophysics, IMAPP, Radboud University Nijmegen, 6500 GL Nijmegen, The Netherlands\\
  $^4$Department of Physics, University of Zagreb, Bijeni\v{c}ka cesta 32, 10000 Zagreb, Croatia\\
  $^5$Institute for Astro- and Particle Physics, University of Innsbruck, Technikerstrasse 25/8, Austria\\
  $^6$Department of Mathematics, Physics \& Geology, Cape Breton University, 1250 Grand Lake Road, Sydney, Nova Scotia, Canada, B1P 6L2\\
  $^7$University of Vienna, Institute of Astronomy, T\"urkenschanzstrasse 17, 1180 Vienna, Austria\\
  $^8$Department of Astronomy and Physics, St. Mary's University, Halifax, NS B3H 3C3, Canada\\
  $^9$D\'epartment de physique, Universit\'e de Montr\'eal, C.P.6128, Succ. Centre-Ville, Montr\'eal, QC H3C 3J7, Canada\\
  $^{10}$NASA-Ames Research Park, MS-244-30, Moffett Field, CA 94035, USA\\
  $^{11}$Department of Astronomy \& Astrophysics, University of Toronto, 50 St. George Street, Toronto, ON M5S 3H4, Canada\\
  $^{12}$Harvard-Smithsonian Center for Astrophysics, 60 Garden Street, Cambridge, MA 02138, USA}

\date{Received date; accepted date}

\pagerange{\pageref{firstpage}--\pageref{lastpage}} \pubyear{2014}

\begin{document}

\label{firstpage}

\maketitle

\begin{abstract}
  Binary stars provide a valuable test of stellar structure and evolution, because the masses of the individual stellar components can be derived with high accuracy and in a model-independent way. In this work, we study
  Spica, an eccentric double-lined spectroscopic binary system
  with a $\beta$\,Cep type variable primary component. We use state-of-the-art modelling tools to
  determine accurate orbital elements of the binary system and atmospheric parameters of both stellar
  components. We interpret the short-period variability intrinsic to the
  primary component, detected on top of the orbital motion both
  in the photometric and spectroscopic data. The non-LTE based
  spectrum analysis reveals two stars of
  similar atmospheric chemical composition consistent with the present day cosmic abundance
  standard defined by \citet{Nieva2012}. The masses and radii of the stars are found to be
  11.43$\pm$1.15~M$_{\odot}$ and 7.21$\pm$0.75~M$_{\odot}$, and
  7.47$\pm$0.54~R$_{\odot}$ and 3.74$\pm$0.53~R$_{\odot}$ for the
  primary and secondary, respectively. We find the primary component to pulsate in three independent
  modes, of which one is identified as a radial mode, while the two others are found to be non-radial, low degree $l$ modes. The frequency of one of these modes is an exact multiple of the orbital frequency, and the $l=m=2$ mode identification suggests a tidal nature for this particular mode. We find a very good agreement between the derived dynamical and evolutionary masses for the Spica system to within the observational errors of the measured masses. The age of the system is estimated to be
  12.5$\pm$1~Myr.
\end{abstract}


\section{Introduction}

Spectroscopic binary stars provide a valuable test of stellar
structure and evolution models since accurate, model-independent
dynamical masses can be measured for the individual stellar
components of a binary. Moreover, an existence of pulsating stars in
binary systems allows the confrontation of the dynamical masses with
those obtained from the asteroseismic analysis. Since the latter
masses are model-dependent, pulsators in binary systems provide a
test of the theory of stellar evolution and oscillations. Apart from
that, massive stars in binary systems are known to suffer from the
so-called mass discrepancy problem, which is important to
investigate in the context of improving the current theories of
stellar structure and evolution. The term mass discrepancy was first
introduced by \citet{Herrero1992} and in its original formulation
refers to the disagreement between the spectroscopic and
evolutionary masses of the star. The evolutionary mass is the one
estimated by fitting evolutionary tracks to a position of the star
in the Hertzspung-Russel or \te-\logg\ Kiel diagram, whereas the
spectroscopic mass stands for the stellar mass computed from the
spectroscopically derived value of the surface gravity and from the
radius of the star. The radius is in turn calculated from the
absolute visual magnitude of the star and the (effective temperature
dependent) integral of stellar flux over the wavelength
\citep[see][]{Kudritzki1980}. Thus, the definition of the mass
discrepancy by \citet[][]{Herrero1992} has nothing to do
specifically with binary systems but refers to massive stars in
general. We keep on using the term mass discrepancy in application
to binary systems throughout the paper, and warn the reader that we
refer to the discrepancy between the evolutionary masses and those
inferred from binary dynamics (dynamical masses).

This paper is the third in a series and concerns a detailed
investigation of Spica, a short-period double-lined spectroscopic
binary containing a massive star, where we address the above
mentioned problems. In the two previous studies, the double-lined
spectroscopic binary systems V380\,Cyg~\citep{Tkachenko2014a} and
$\sigma$\,Scorpii~\citep{Tkachenko2014b} have been investigated. The
asteroseismic modelling of the primary component of $\sigma$\,Sco
was largely successful leading to the mass and radius determination
to precisions of ~10\% or better, which is a factor of three more
accurate than the corresponding values derived from the combined
spectroscopic orbital and atmospheric parameters, and the
interferometric orbital inclination angle \citep[taken
from][]{North2007}. We have found an evidence of the mass
discrepancy for the secondary component of the binary, i.e. the
evolutionary models suggest higher mass than the one inferred from
the binary dynamics. The study of V380\,Cyg benefited from the fact
that it is an eclipsing system with space-based photometry
available, thus the dynamical masses and radii of the individual
stellar components could be derived to precisions approaching 1\%
level. Both components of the binary were found to show significant
discrepancy between the dynamical and evolutionary masses, leading
us to the conclusion that the single star evolutionary models are
inadequate for this particular system.

\begin{figure*}
\includegraphics[scale=0.85]{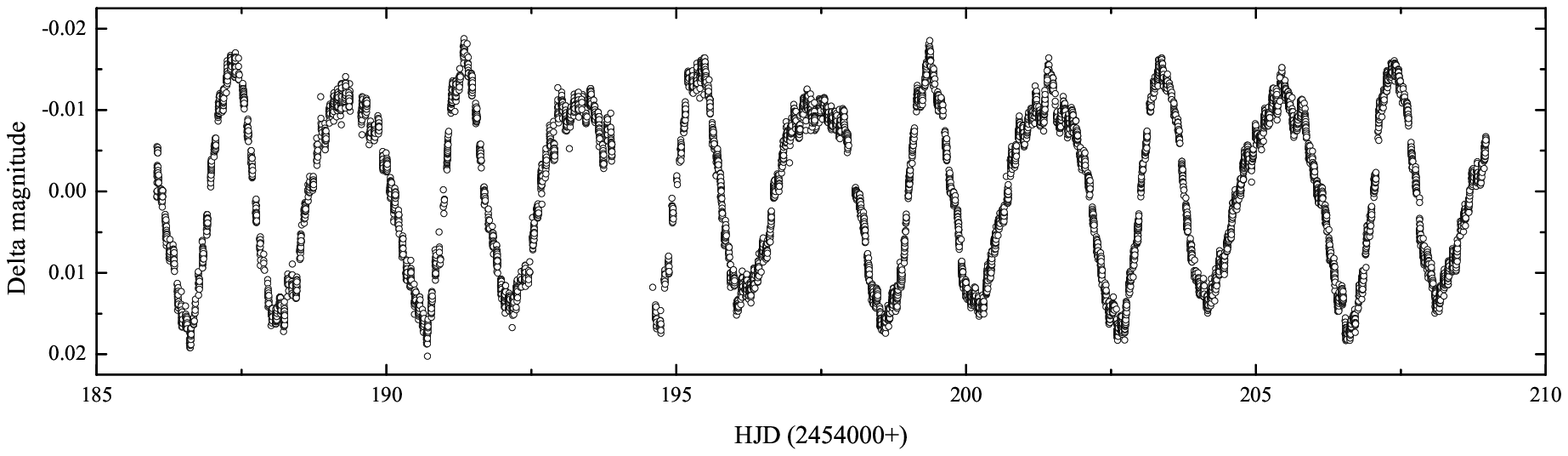}\vspace{5mm}
\includegraphics[scale=0.85]{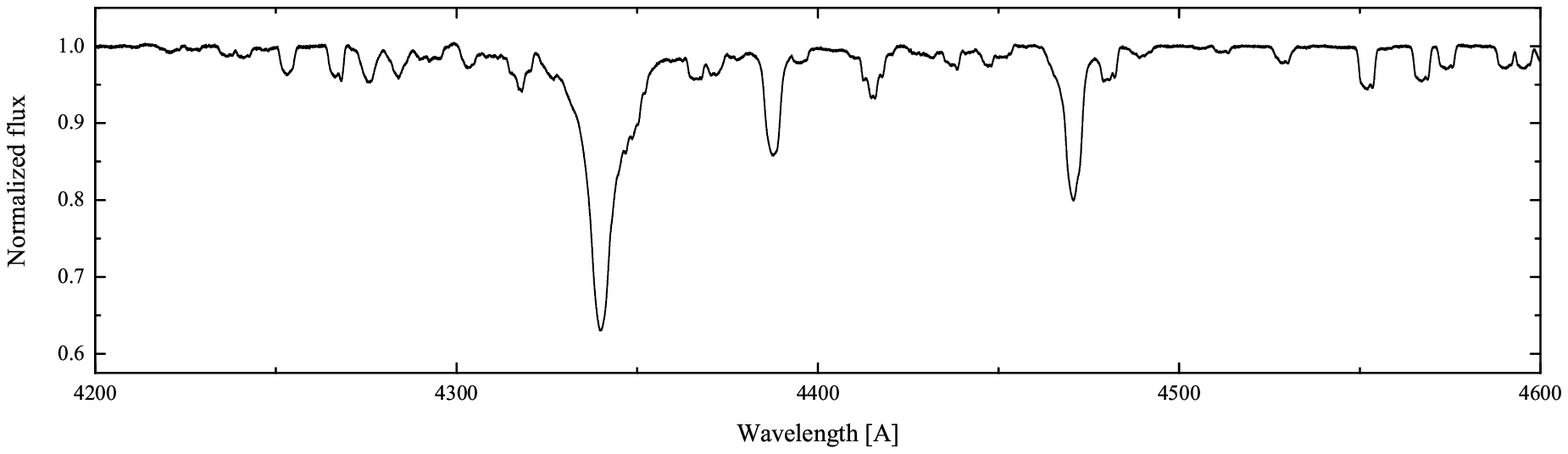}
\caption{The MOST light curve (top) and a portion of the normalized,
composite {\sc coralie} spectrum (bottom) of Spica.} \label{Figure1}
\end{figure*}

Spica \citep[$V=0.97$~mag, discovered to be a spectroscopic binary
by][]{Vogel1890} is an important system to study in the context of
testing the current theories of stellar structure and evolution for
massive stars, which are way more uncertain than the models for
lower mass stars. The more evolved primary component is also known
to be an intrinsically variable star that belongs to the class of
$\beta$~Cep variables \citep[][Chapter 2]{Aerts2010} and pulsates in
pressure modes. As a consequence, the system was the subject of
numerous studies addressing both the questions on configuration of
the orbit as well as on the variability intrinsic to the primary
component. For example, \citet{Shobbrook1972} used both the newly
obtained and historical RV data to determine orbital parameters of
the binary. The obtained solution was found to be in good agreement
with the one reported by \citet{Herbison-Evans1971}.

\begin{table}
\centering \tabcolsep 2.5mm\caption{List of the spectroscopic
observations of Spica. JD is the Julian Date, $N$ gives the number
of spectra taken during the corresponding observational
period.}\label{Table1}
\begin{tabular}{lcr} \hline
\multicolumn{2}{c}{Time period} & $N$\\
\multicolumn{1}{c}{Calendar date} & \multicolumn{1}{c}{JD
(2\,454\,000+)} &\\\hline
19.03--01.04.2007 & 178--191 & 772\\
18.04--30.04.2007 & 208--220 & 798\\
22.07--30.07.2007 & 304--312 & 161\\\cline{3-3}
\multicolumn{2}{l}{Total number of spectra:\rule{0pt}{11pt}} & {\bf
1731}\\ \hline
\end{tabular}
\end{table}

Despite the extensive investigations in the past, the oscillation
behaviour of Spica is currently not well understood: several studies
reported about irregular variability of the individual frequencies
on short time scales, and there is quite some diversity in the
reported frequencies the Spica system was found to pulsate in (see
Sect.~\ref{Sect: frequency analysis} for an overview). The fact that
the system is eccentric and the intrinsic pulsations of the more
evolved component are in the range below 6~\cd\ (69.4~\mhz) makes it
an interesting target to search for pulsations that are resonantly
driven by the dynamic tide. These so-called tidally induced
pulsations \citep[][Chapter 2]{Aerts2010} have so far been detected
in BA main-sequence stars with masses below 5~$M_{\odot}$
\citep[e.g.,][]{Welsh2011,Thompson2012,Hambleton2013,Papics2013} but
never in early B-type stars. Should this type of stellar oscillation
be found in either of the stellar components of the Spica system, it
will be the first massive binary in which resonantly driven
oscillations have been detected. Such a detection would make Spica
an important test case for the evolution of massive stars in binary
systems as the tidally-induced pulsations are expected to play a
significant role in angular momentum exchange between the orbit and
the individual stellar components, and thus in the evolution of the
orbital elements of the system with time.

Better observational constraints on the oscillation behaviour of
Spica could be obtained if high-quality space-based were available
for this star. Moreover, the system was also reported by
\citet{Desmet2009} as possibly a grazingly eclipsing binary, so good
quality observational data coupled to state-of-the-art analysis are
required to detect a tiny eclipse signal, if present at all. A
possible detection of the eclipses in the photometric data of the
system could help to constrain the fundamental properties of the
individual stellar parameters to a better precision than is
currently available.

Overall, our study addresses several scientific questions at the
same time. First of all, we aim at improving the understanding of
the pulsational behaviour of the Spica system using and improved
observational dataset, including high-quality photometric
observations from space combined with high-quality high-resolution
ground-based spectroscopy (see Sect.\ref{Sect.:observations}). The
orbital configuration of the system is determined in
Sects.~\ref{Sect: phoebe} and \ref{Sect: SPD}, with a particular
interest in whether the star could be a grazingly eclipsing system.
We also perform a detailed (atmospheric) chemical composition
analysis for both stellar components of the binary (Sect.~\ref{Sect:
spectrum analysis}), which particularly allows us to see whether a
increased nitrogen abundance at the surface of the star due to
rotational mixing applies to the Spica system as reported in the
literature \citep[e.g.,][]{Hunter2008,Brott2011}. The intrinsic
variability of the more massive component of the Spica system is
explored in detail in Sect.~\ref{Sect: frequency analysis}. This
includes detailed frequency analysis, identification of the
individual pulsation modes from spectroscopic data, and a search for
observational signatures of tidally-induced pulsations. The
possibility of detecting for the first time forced oscillations in a
high-mass star ($>5$M$_{\odot}$) defines the second scientific
question we address in this study. Finally, we investigate the
binary system in the context of the above-mentioned mass discrepancy
problem by comparing the (spectroscopically inferred) positions of
the individual stellar components in the \te-\logg\ Kiel diagram
with the evolutionary tracks (Sect.~\ref{Sect: evolutionary
models}).

\section{Observations and data reduction}\label{Sect.:observations}

We obtained 1731 high resolution (R = 50\,000) high signal-to-noise
ratio (S/N) spectra with the {\sc coralie} spectrograph attached to
the 1.2m Euler Swiss Telescope (La Silla, Chile). The data were
gathered in three different runs, in March, April, and July 2007,
and cover wide wavelength range between 381 and 681 nm.
Table~\ref{Table1} gives the journal of our observations and lists
calendar and Julian dates of the beginning and the end of the
observing run, as well as the number of spectra that were obtained
within each of the runs.

The spectroscopic data were reduced by means of a dedicated data
reduction pipeline. The procedure included bias and stray-light
subtraction, cosmic rays filtering, flat fielding, wavelength
calibration by ThAr lamp, and order merging. Normalisation to the
local continuum has been done by fitting a cubic spline function
through some tens of carefully selected continuum points. The whole
normalisation procedure as well as its implementation are described
in more detail in \citet{Papics2012}.

Photometric data were gathered with the MOST space-based mission
\citep{Walker2003}. The satellite is equipped with a 15-cm
Rumak-Maksutov optical telescope feeding a CCD photometer. The Spica
system was observed in the ``Fabry mode'' of MOST
\citep{Walker2003}, and was sampled using $\sim$30 second exposures
over a baseline of 22.92~days in March-April 2007. The single point
precision is of the order of 1--2~mmag; the complete light curve of
the system is illustrated in Fig.~\ref{Figure1} (top panel). For the
analysis, the light curve was converted to a magnitude scale; a
small, long term instrumental trend has been removed by fitting a
second-order degree polynomial to the entire data set.

The observational data set we analyse in this study is unique in the
sense that 1) photometric data were obtained for the Spica system
from space for the first time and are expected to provide much
better constraints on the orbital configuration of the system and on
the intrinsic variability of the individual stellar components; 2)
the photometric data are complemented with extensive time-series of
ground-based spectroscopic observations that were taken during the
same period of time. Detailed (combined) analysis of these data sets
allows us to cross-match the photometrically inferred oscillation
frequencies with those extracted from the spectroscopic data. The
frequencies found in both data sets can be safely considered as
intrinsic to the star and used for identification of the individual
pulsation mode geometries based on well-sampled spectroscopic data.
The output of such an analysis is one of the main pre-conditions for
a detailed asteroseismic modelling of the star.

\section{Modelling the light- and RV-curve data}\label{Sect: phoebe}

We used the method described in \citet{Papics2013} to determine the
RVs of both components from the composite observed spectra. Briefly,
the method relies on the fitting of composite synthetic spectra to
the observations, and comprises of two basic steps. First, a rough
estimate of the spectroscopic parameters is made by fitting
synthetic spectra to a single, high-quality observation taken at the
orbital phase of largest RV separation, in a grid of \te, \logg,
\vsini, and [M/H], in the range of several helium and metallic
lines, as well as in the regions of Balmer H$_{\beta}$ and
H$_{\gamma}$ line profiles. In the second step, the atmospheric
parameters are fixed to the values derived in the previous step, and
RVs are determined by fitting composite synthetic spectra to all
observed spectra leaving RVs of both components as variables. In our
case, we could skip the first (rather time-consuming) step, and we
used the parameters \te=24\,700/20\,800, \logg=3.7/4.2,
\vsini=161/70, and [M/H]=0.0/0.0 reported by \citet{Lyubimkov1995}
for the primary/secondary component.

As pointed out by \citet{Harrington2009}, the determination of the
RVs of both binary components is complicated due to non-negligible
line-profile variability (LPV) intrinsic to the primary component.
In the result, the individual RVs of the primary measured as the
first order moment of the line profile may differ by up to 16~\kms\
from the average value, depending on the range of the profile that
is used for the measurement \citep[see Fig.3 in][]{Harrington2009}.
Moreover, at the phases of close RV separation of the components, an
additional uncertainty is added by the secondary to the RVs of the
primary, but also the RVs of the secondary are affected by the LPV
of the primary component. Both effects were clearly visible in our
RVs, measured with the procedure outlined above. Thus, we used these
RVs to provide initial guesses for the orbital parameters, but the
final orbital solution was obtained by iterating between the light
curve fitting (using the most recent version of the {\sc phoebe2.0}
code\footnote{http://phoebe-project.org/},
\citet{Degroote2013,Prsa2013}, see below) and the method of spectral
disentangling ({\sc spd}, see Sect.\ref{Sect: SPD}). Such approach
has an advantage in that it allows to (at least partially) remove
the degeneracy between the longitude of periastron ($\omega$) and
the eccentricity ($e$) of the system. In practice, the former
parameter is better constrained from the light curve, whereas the
eccentricity is better defined from the shape of the RV-curve of the
system. This is exactly the reason why the eccentricity reported in
Table~\ref{Table2} is mentioned as ``adopted from spectroscopy'' but
the $\omega$ value given in Table~\ref{Table3} is the one that comes
from the light curve solution.

\begin{table}
\caption{Summary of the parameters for the final {\sc phoebe~2.0}
solution of the MOST light- and {\sc coralie} radial velocity curves
of Spica. A and B refer to the primary and secondary stars,
respectively. The uncertainties on the parameters come from Markov
Chain Monte Carlo simulations.} \label{Table2}
\begin{tabular}{llc} \hline
Parameter                           & Value     & Uncertainty            \\
\hline
Orbital period, P (d)               & 4.0145    & 0.0001 \\
T$_0$ (HJD 24\,54\,000.+)        & 189.40   & 0.02 \\
Orbital eccentricity$^*$, $e$             & 0.133         & 0.017           \\
Longitude of periastron, $\omega$ (\degr)     & 255.6           &     12.2      \\
Mass ratio$^*$, $q$                       & 0.6307    & 0.023 \\
Orbital inclination, $i$ (\degr)      & 63.1      & 2.5    \\
Orbital semimajor axis ($R_\odot$) & 28.20 & 0.92 \\
Systemic velocity, $\gamma$ (\kms) & -0.5 & 1.1\\
Rotation rates, A, B                      & 2.0, 1.5 &         \\
Gravity darkening, A, B                   & 1.0, 1.0 &              \\
Third light                         & 0.0 &              \\
Star\,A potential                   & 4.67                  & 0.21     \\
Star\,B potential                   & 6.08                  &0.50      \\
Fractional polar radii, A, B       & 0.250, 0.132& 0.08, 0.05\\
Fractional equatorial radii, A, B       & 0.282, 0.134& 0.09, 0.04\\
$T_{\rm eff}$ star A$^*$ (K)           & 25\,300  & 500        \\
$T_{\rm eff}$ star B (K)           & 20\,585               &     850       \\
\hline \multicolumn{3}{l}{$^*$ adopted from spectroscopy}
\end{tabular}
\end{table}

The variability intrinsic to the primary component is also
pronounced in the MOST light curve. To minimize the effect of this
variability on the {\sc phoebe2.0} solution, the data were binned
into 200 orbital phase bins of equal width. This number of phase
bins was found to be optimal in the sense that it provided good
enough sampling of the light curve and allowed to suppress the
amplitude of the oscillations significantly.

A detached configuration of the system has been assumed in our
calculations. Given the previously reported values for \vsini\ and
radii of both components, as well as of the orbital inclination
angle \citep[e.g.,][]{Herbison-Evans1971,Lyubimkov1995}, the
synchronicity parameter was set to 2.0 and 1.5 for the primary and
secondary star, respectively. Our final solution will show that this
assumption is very close to reality. The gravity darkening
coefficients were set to unity and the third light was fixed at
zero. The effect of limb darkening is taken into account by means of
the non-linear low introduced by \citet{Claret2000}. Unlike, e.g.,
the Wilson-Devinney approach where one, global set of coefficients
is assumed to be valid on the entire star, {\sc phoebe2.0} takes
into account the dependency of the coefficients on local conditions
(effective temperature, gravity) on the star. This way,  the limb
darkening coefficients are not decoupled from the intensities
anymore but are assumed to vary locally over the surface
\citep{Degroote2013}.

\begin{figure}
\includegraphics[scale=0.55]{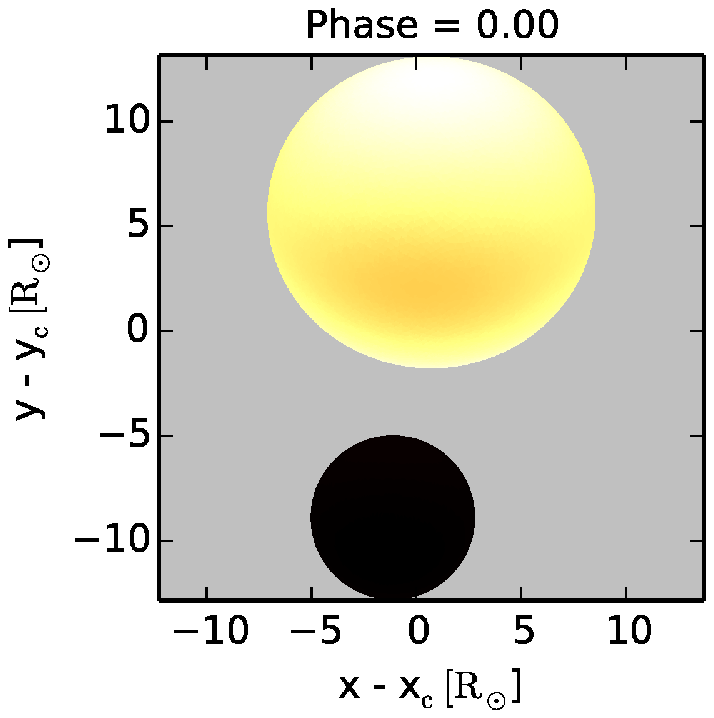}
\includegraphics[scale=0.55]{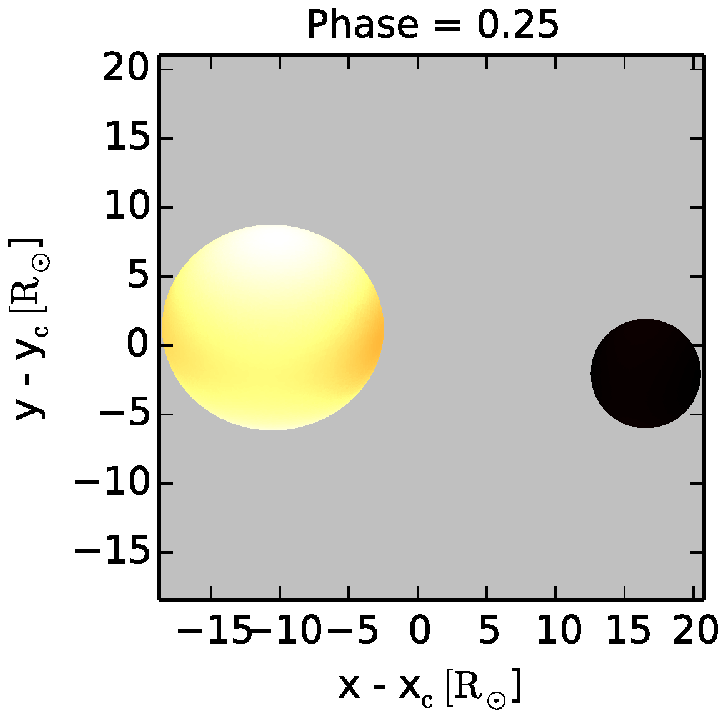}\vspace{3mm}
\includegraphics[scale=0.55]{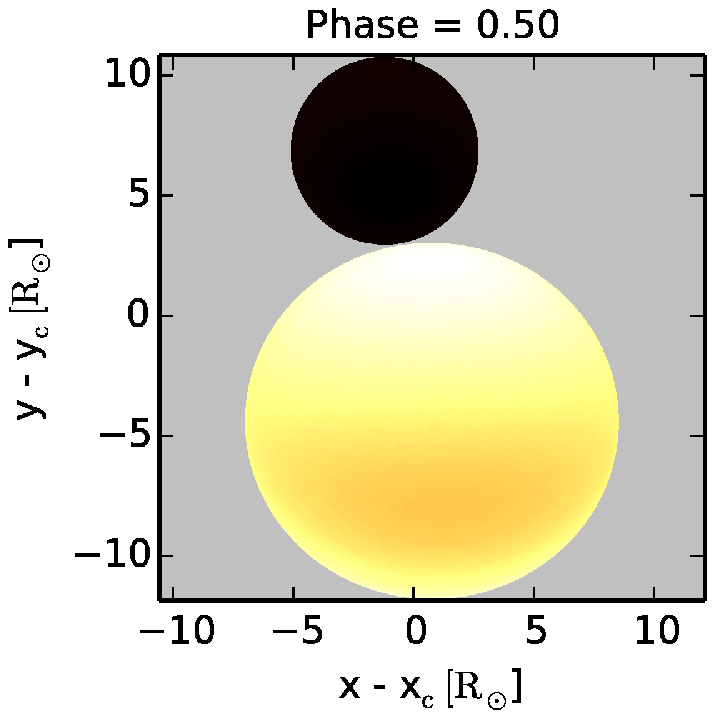}\hspace{5mm}
\includegraphics[scale=0.55]{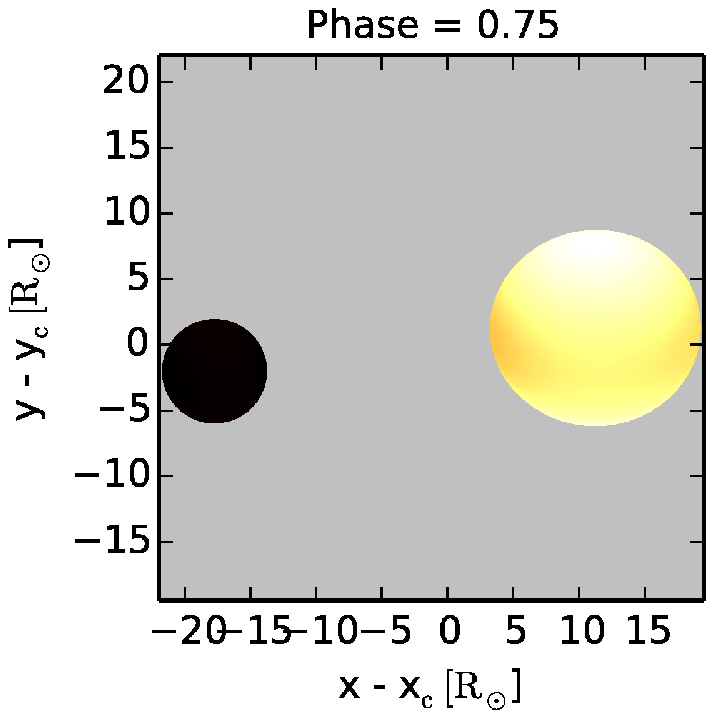}
\caption{Schematic representation of the Spica binary system at four
different orbital phases. Phase zero corresponds to the time of
periastron passage. X and Y axes give the distance from the center
of mass in units of solar radii.} \label{Figure2}
\end{figure}

\begin{figure}
\includegraphics[scale=0.85]{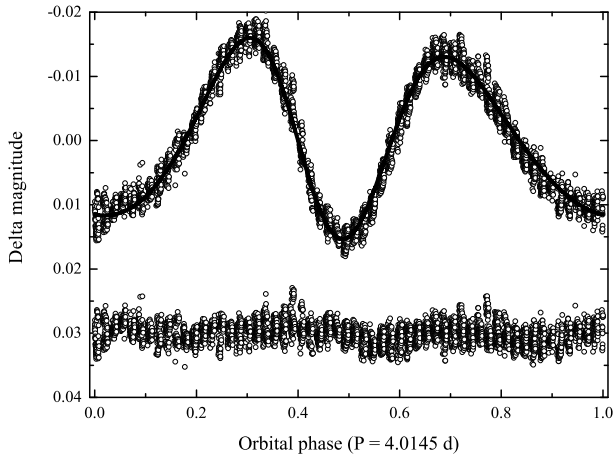}
\caption{Best fit model (solid, thick line) to the phase-folded MOST
light curve (open circles). The residuals have been shifted
downwards by 0.03~mag for clarity.} \label{Figure3}
\end{figure}

The time of periastron passage (T$_{0}$), the eccentricity ($e$),
the orbital inclination angle ($i$), the longitude of periastron
($\omega$), the semi-major axis ($a$), the systemic velocity
($\gamma$), the dimensionless potentials ($\Omega_{1, 2}$), and the
effective temperature of the secondary ($T_{\rm eff, 2}$) were set
as adjustable parameters. The effective temperature of the primary
($T_{\rm eff, 1}$) was initially fixed to the value of 24\,700~K as
derived by \citet{Lyubimkov1995}, while the final solution was
computed assuming our spectroscopically derived value of 25\,300~K
(see Sect.~5). Fixing the effective temperature of one of the stars
is essential as only the temperature ratio of the two components can
be constrained from white light photometry delivered by MOST. The
final set of the parameters is given in Table~\ref{Table2}, along
with the uncertainties derived by means of the Markov Chain Monte
Carlo simulations. We also provide fractional radii (stellar radii
divided by the orbital semi-major axis) which are needed to compute
the physical properties of the two stars. The system configuration
is represented schematically at four different orbital phases in
Fig.~\ref{Figure2}, while the quality of the fit of the phase-folded
MOST light curve is illustrated in Fig.~\ref{Figure3}. As one can
see (bottom left panel in Fig.~\ref{Figure2}), the orbital
inclination of Spica is such that it appears to be a nearly
eclipsing system. Provided the error bars on the potentials of the
two stars and the orbital inclination angle given in
Table~\ref{Table2}, Spica might indeed be a grazingly eclipsing
binary system, where the more evolved primary partially blocks the
light of the secondary component near phase 0.5, as suggested by
\citet{Desmet2009}. Though the discovery of eclipses in this system
would be an exciting result, it would hardly help better constrain
orbital and physical parameters of the system.

\begin{figure}
\includegraphics[scale=0.85]{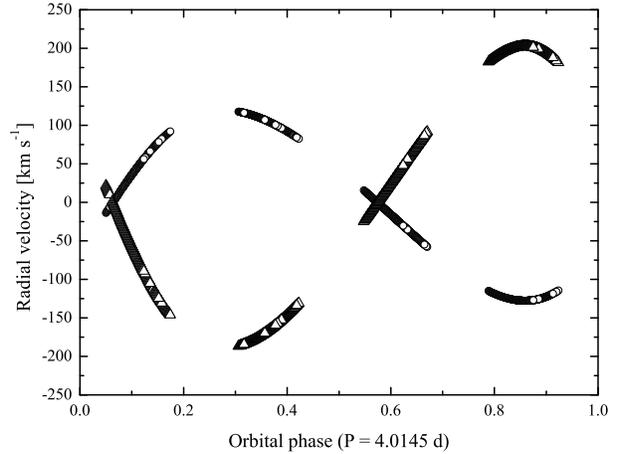}
\caption{RVs of both binary components computed from the orbital
solution reported in Table~\ref{Table3}. The primary component is
shown by open circles, whereas the secondary with open triangles.}
\label{Figure4}
\end{figure}

A clear residual signal can be seen on top of the one due to the
ellipsoidal variability of stars in the MOST light curve of Spica
(see Fig.~\ref{Figure3}). This signal is clearly intrinsic to the
star and will be discussed in Sect.~\ref{Sect: frequency analysis}
in more detail. We did not use the residual RVs obtained at this
step for further analysis, as the more accurate spectroscopic
information can be extracted from the original, composite spectra of
a binary by subtracting the contribution of either of the binary
components beforehand, using the corresponding disentangled spectrum
(see Sect.~\ref{Sect: frequency analysis}).

\section{Spectral disentangling}\label{Sect: SPD}

To separate the spectra of two components, we used the method of
spectral disentangling ({\sc spd}, hereafter) originally proposed by
\citet{Simon1994}, and formulated by \citet{Hadrava1995} in Fourier
space. The code we use is the the {\sc FDbinary} code by
\citet{Ilijic2004}. The method is designed to optimize
simultaneously for the spectra of individual binary components and
orbital elements of the system. The procedure we used here is
similar to the one described in
\citet{Tkachenko2014a,Tkachenko2014b}, that is: in the first step,
the orbital elements were derived from several spectral regions
centered at prominent lines of neutral helium and doubly-ionized
silicon (He~{\small I} $\lambda\lambda$ 4388, 4471, 4920, 5015~\AA\
and Si~{\small III} triplet $\lambda\lambda$ 4553, 4568, and
4575~\AA), and in the second step, the obtained orbital solution was
used to separate the spectra of individual binary components in the
entire wavelength range. In this particular case, the orbital
elements used for the spectral separation are the ones obtained from
the combined spectroscopic and photometric analysis (see
Table~\ref{Table3}). The wavelength range of the separated spectra
covers several Balmer lines (H$_{\delta}$, H$_{\gamma}$, and
H$_{\beta}$) and extends up 5\,300~\AA. The whole red part of the
spectrum was avoided in the disentangling because of numerous
telluric contributions and a few stellar lines only present in that
part of the spectrum. Instead, {\sc spd} was performed on individual
stellar lines in the red part of the spectrum, that were
advantageous to have for the analysis of chemical composition of the
components (e.g., Si~{\small III} $\lambda\lambda$ 5741~\AA\ line
and a few nitrogen, carbon, and aluminium lines nearby). The blue
part of the spectrum was avoided because of uncertain continuum
normalization. So, the spectral window that was finally used for the
spectrum analysis covers the wavelength range between 4\,000 and
5\,300~\AA. The final orbital solution was obtained by computing the
mean orbital elements from all wavelength intervals mentioned above;
the values are reported and compared to the previous findings in
Table~\ref{Table3}. All parameters are found to agree within the
quoted error bars with the values reported by
\citet{Herbison-Evans1971} and \citet{Shobbrook1972}. Using the
value of T$_0$ reported by \citet{Shobbrook1972} and the apsidal
motion period $P_{\rm aps}=139\pm7$ years derived by
\citet{Aufdenberg2007}, we calculate $\omega$ to be between
230$^{\circ}$ and 255$^{\circ}$ for the period of our observations.
Thus, the obtained by us value for the longitude of periastron of
255.6$\pm$12.2$^{\circ}$ is in agreement with $P_{\rm aps}$ reported
by \citet{Aufdenberg2007}. Figure~\ref{Figure4} illustrates the
final solution in terms of the computed RVs of both components, and,
at the same time, gives an impression about the orbital phase
coverage achieved with our spectroscopic material.

\begin{table}
\tabcolsep 1.1mm\caption{Orbital elements for Spica as derived by
\citet[][indicated as H-E1971]{Herbison-Evans1971},
\citet[][indicated as S1972]{Shobbrook1972}, and in this study.
Errors are given in parentheses in terms of last digits. Subscripts
1 and 2 refer to the primary and secondary component,
respectively.}\label{Table3}
\begin{tabular}{lllll} \hline
Param. & Unit & H-E1971 & S1972 & this study\\\hline
$P$ & days & 4.01455(3) & 4.01454(3) & 4.0145(1)$^{**}$\\
$K_1$ & \kms & 123.9(1.4) & 124(4) & 123.7(1.6)\\
$K_2$ & \kms & 198.8(1.5) & 197(8) & 196.1(1.6)\\
$e$ & & 0.146(9) & 0.14(3) & 0.133(17)\\
$\omega^*$ & degrees, $^{\circ}$ & 138(6) & 142(8) & 255.6(12.2)$^{**}$\\
T$_0$ & HJD & 2440678.09(7) & 2440284.76(8) & 2454189.40(2)$^{**}$\\
\hline\multicolumn{5}{l}{$^*$ The values measured at T$_0$}\\
\multicolumn{5}{l}{$^{**}$ Fixed from the {\sc phoebe2.0} solution
(see Sect.~\ref{Sect: phoebe})}
\end{tabular}
\end{table}

Since the primary of Spica is known to be a variable star of
$\beta$~Cep type, and the {\sc spd} method assumes no variability
intrinsic to either of the components of a binary system, the input
data used for the spectral disentangling need some additional
commenting. We used the approach adopted in \citet{Tkachenko2014b},
that is we binned the original spectra with orbital period derived
in Sect.~\ref{Sect: phoebe}. From a set of experiments, we found 25
orbital phase bins to be the optimal choice, which provided a
minimum number of orbital phase gaps, and sufficiently uniform
distribution of the spectra over the bins. There are several
advantages of using binned data instead of the original spectra,
provided a large number of spectra is available: i) a significantly
reduced number of input spectra for {\sc spd} (25 compared to the
original 1731 measurements), which in turn reduces the computation
time; ii) a much higher S/N for each spectrum; and iii) the
oscillation signal is suppressed to a level that it has only a minor
impact on the results of the disentangling.

\begin{figure*}
\includegraphics[scale=0.95]{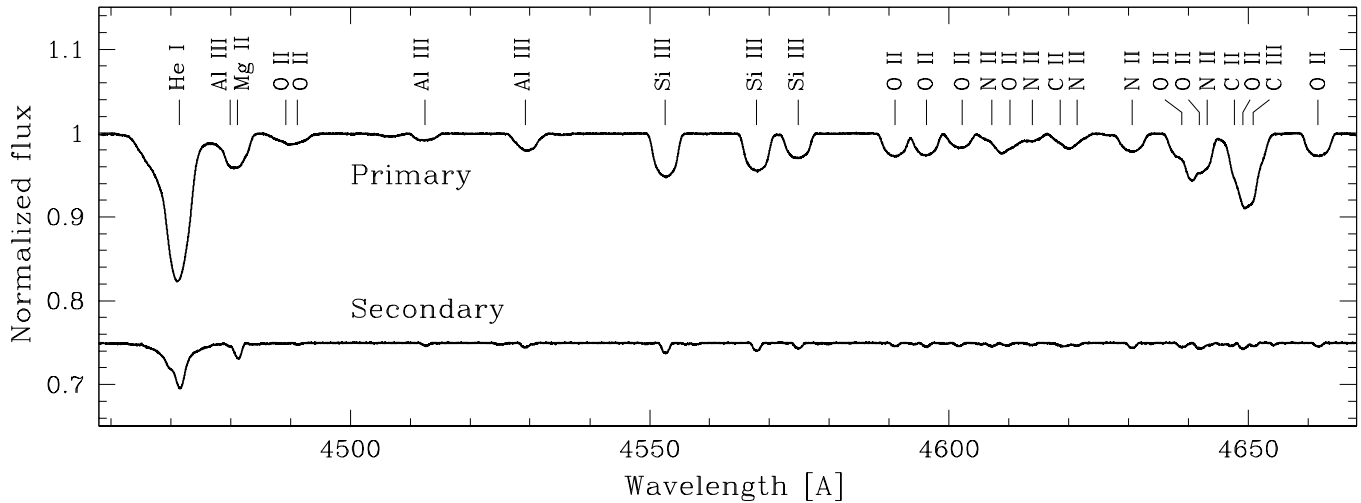}
\caption{A portion of the separated (non-corrected for the
individual light contributions) spectra of both stellar components
of the Spica binary system. The spectrum of the secondary was
shifted downwards by 0.25 (in continuum units) for clarity. The most
prominent lines present in the spectra of both stars are also
indicated.} \label{Figure5}
\end{figure*}

A portion of the separated spectra of both components is shown in
Fig.~\ref{Figure5}, along with identification of most prominant
spectral lines found in the spectra of both components. The spectra
suggest that both components are B-type stars with the primary
component rotating significantly faster than the secondary. A
detailed spectrum analysis of both binary components is presented in
the next section.

\section{Spectrum analysis of binary components}\label{Sect: spectrum
analysis}

Eclipsing binaries are a prime source of accurate masses and radii
of stars, given high-quality photometric and spectroscopic data can
be obtained for them. Moreover, the atmospheric parameters and
chemical composition of individual stellar components of such
systems can often be measured to a higher precision than for single
stars of similar spectral type and luminosity class \citep[see,
e.g.,][]{Pavlovski2005,Pavlovski_Southworth2009,Tkachenko2009,Tkachenko2010,Tkachenko2014a}.
This becomes possible due to the fact that the surface gravity of a
stellar component can be computed directly from its mass and radius,
and thus fixed in the spectroscopic analysis. This removes the
degeneracy between \te\ and \logg\ from which the spectroscopic
analysis of OB-stars is known to suffer, leading to the measurement
of accurate individual abundances. This is not the case for
non-eclipsing systems, where the precision achieved on the
fundamental atmospheric parameters and chemical composition of the
components is comparable to the one expected for single stars.
However, spectroscopic analysis still is superior to the analysis of
broad-band photometry, in the sense that the atmospheric parameters
measured from stellar spectra are more accurate than the values
inferred from photometric data.

\begin{table}
\tabcolsep 2.0mm\caption{Atmospheric parameters of both components
of the Spica binary system. Equatorial velocities were computed
under the assumption of coplanarity.}\label{Table4}
\begin{tabular}{llr@{$\pm$}lr@{$\pm$}l} \hline
Parameter & Unit & \multicolumn{2}{c}{Primary} &
\multicolumn{2}{c}{Secondary}\\
\hline
\te\rule{0pt}{9pt} & K & 25\,300&500 & 20\,900&800\\
\logg & dex & 3.71&0.10 & 4.15&0.15\\
\vsini & \kms & 165.3&4.5 & 58.8&1.5\\
\vmicro & \kms & 5.0&1.0 & 3.0&1.0\\
light factor & \% & 85.5&1.5 & 14.5&1.5\\
 $v_{\rm eq}$ & \kms & 185.4&9.2 & 65.9&3.2\\\hline
\end{tabular}
\end{table}

\begin{table} \centering \caption{\label{tab:abuall} Photospheric abundances
derived for both stellar components of the Spica binary system.
Abundances are expressed relative to the abundance of hydrogen,
$\log \epsilon(H) = 12.0$. The third and the fourth columns give the
number of lines used and the difference between abundances of a
binary component and those of the Sun, respectively. Present-day
cosmic abundances from Galactic OB stars \citep{Nieva2012} are given
in the fifth column. The last column lists the solar abundances from
\citet{Asplund2009}.}\label{Table5}
\begin{tabular}{lcrcccc} \hline
El.\  & $\log \epsilon({\rm X})$ & $N$ & [X/H] & OB stars      & Sun         \\
\hline \multicolumn{6}{c}{{\bf Primary component}}\\
He\rule{0pt}{11pt} & 10.97$\pm$0.06 & 6  &  0.00$\pm$0.06  & 10.99$\pm$0.01  & 10.97$\pm$0.01\\
C  & 8.19$\pm$0.05  & 4  & -0.20$\pm$0.07  & 8.33$\pm$0.04   & 8.39$\pm$0.05 \\
N  & 7.76$\pm$0.04  & 17 & -0.02$\pm$0.07  & 7.79$\pm$0.04   & 7.78$\pm$0.06 \\
O  & 8.72$\pm$0.08  & 8  &  0.06$\pm$0.09  & 8.76$\pm$0.05   & 8.66$\pm$0.05 \\
Mg & 7.50$\pm$0.10  & 1  & -0.03$\pm$0.13  & 7.56$\pm$0.05   & 7.53$\pm$0.09 \\
Si & 7.41$\pm$0.20  & 6  & -0.10$\pm$0.20  & 7.50$\pm$0.05   & 7.51$\pm$0.04 \\
Al & 6.20$\pm$0.08  & 2  & -0.17$\pm$0.09  &                 & 6.37$\pm$0.04 \\
\hline\multicolumn{6}{c}{{\bf Secondary component}}\\
He\rule{0pt}{11pt} & 10.98$\pm$0.07 & 6  &  0.01$\pm$0.07  & 10.99$\pm$0.01  & 10.97$\pm$0.01\\
C  & 8.26$\pm$0.14  & 4  & -0.13$\pm$0.15  & 8.33$\pm$0.04   & 8.39$\pm$0.05 \\
N  & 7.81$\pm$0.19  & 17 &  0.03$\pm$0.20  & 7.79$\pm$0.04   & 7.78$\pm$0.06 \\
O  & 8.82$\pm$0.23  & 8  &  0.16$\pm$0.24  & 8.76$\pm$0.05   & 8.66$\pm$0.05 \\
Mg & 7.45$\pm$0.16  & 1  & -0.08$\pm$0.18  & 7.56$\pm$0.05   & 7.53$\pm$0.09 \\
Si & 7.52$\pm$0.24  & 6  &  0.01$\pm$0.24  & 7.50$\pm$0.05   & 7.51$\pm$0.04 \\
Al & 6.16$\pm$0.17  & 2  & -0.20$\pm$0.17  &                 & 6.37$\pm$0.04 \\
\hline
\end{tabular}
\end{table}

\begin{figure*}
\includegraphics[scale=0.27]{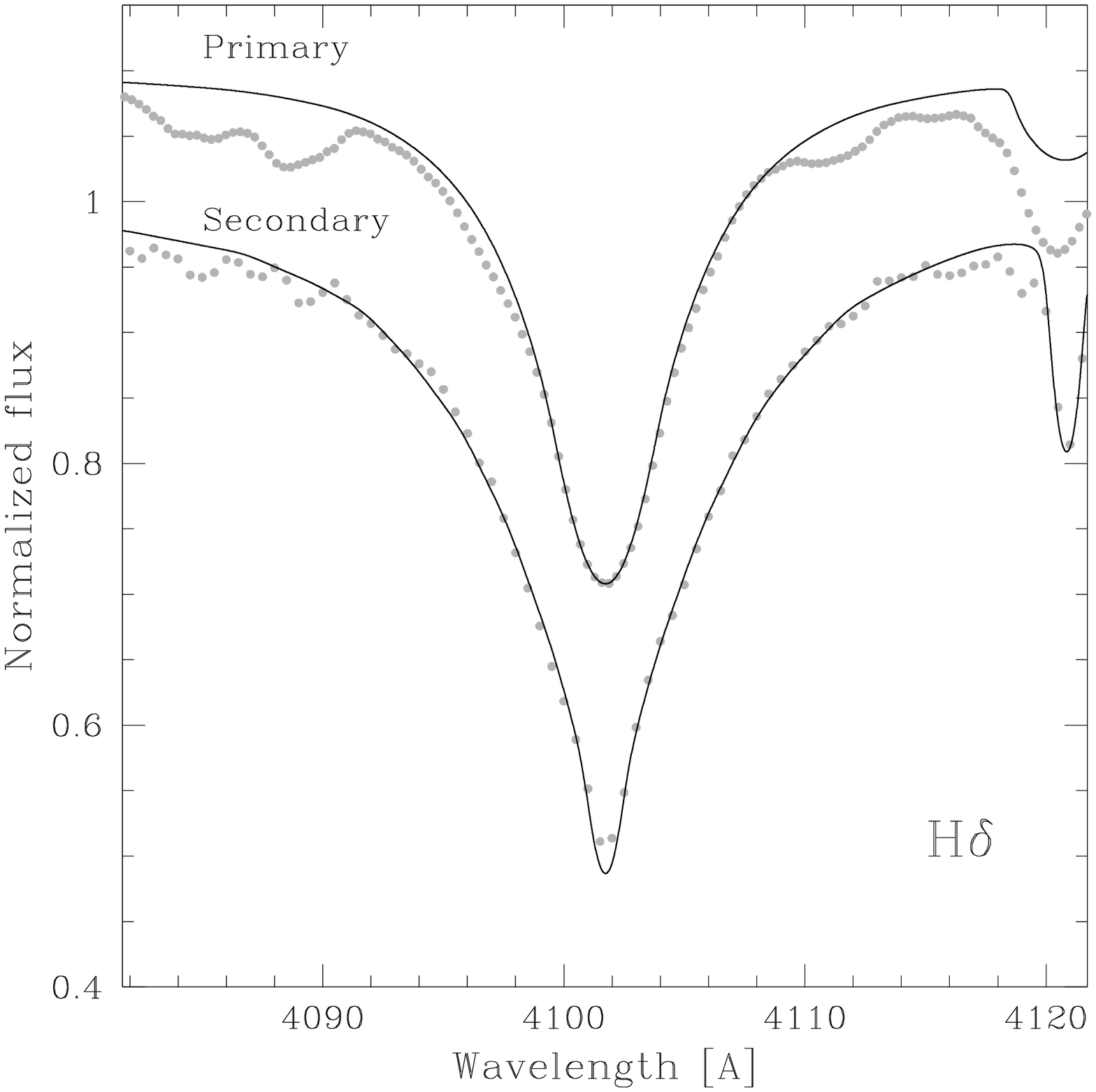}\hspace{5mm}
\includegraphics[scale=0.27]{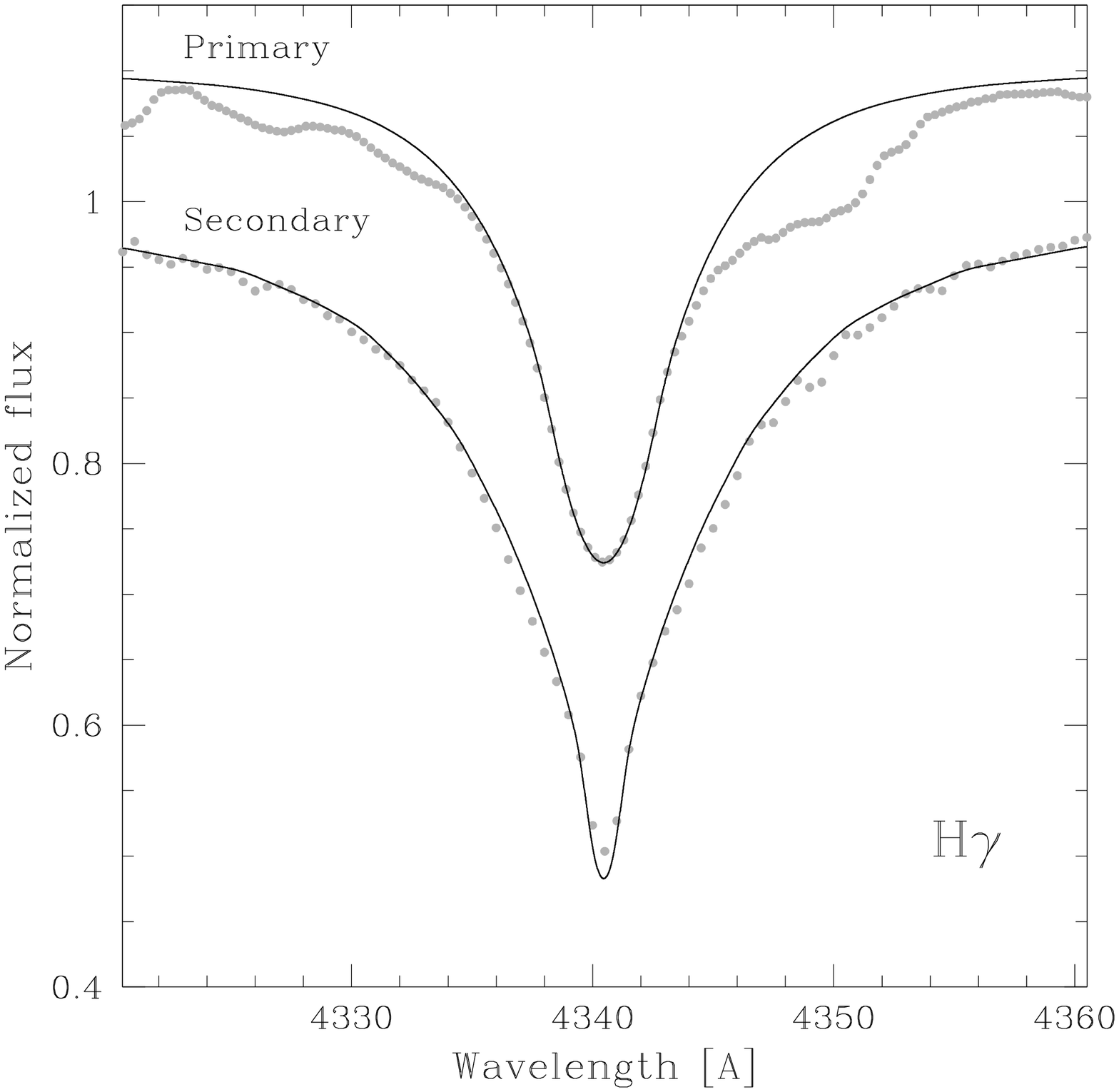}\hspace{5mm}
\includegraphics[scale=0.27]{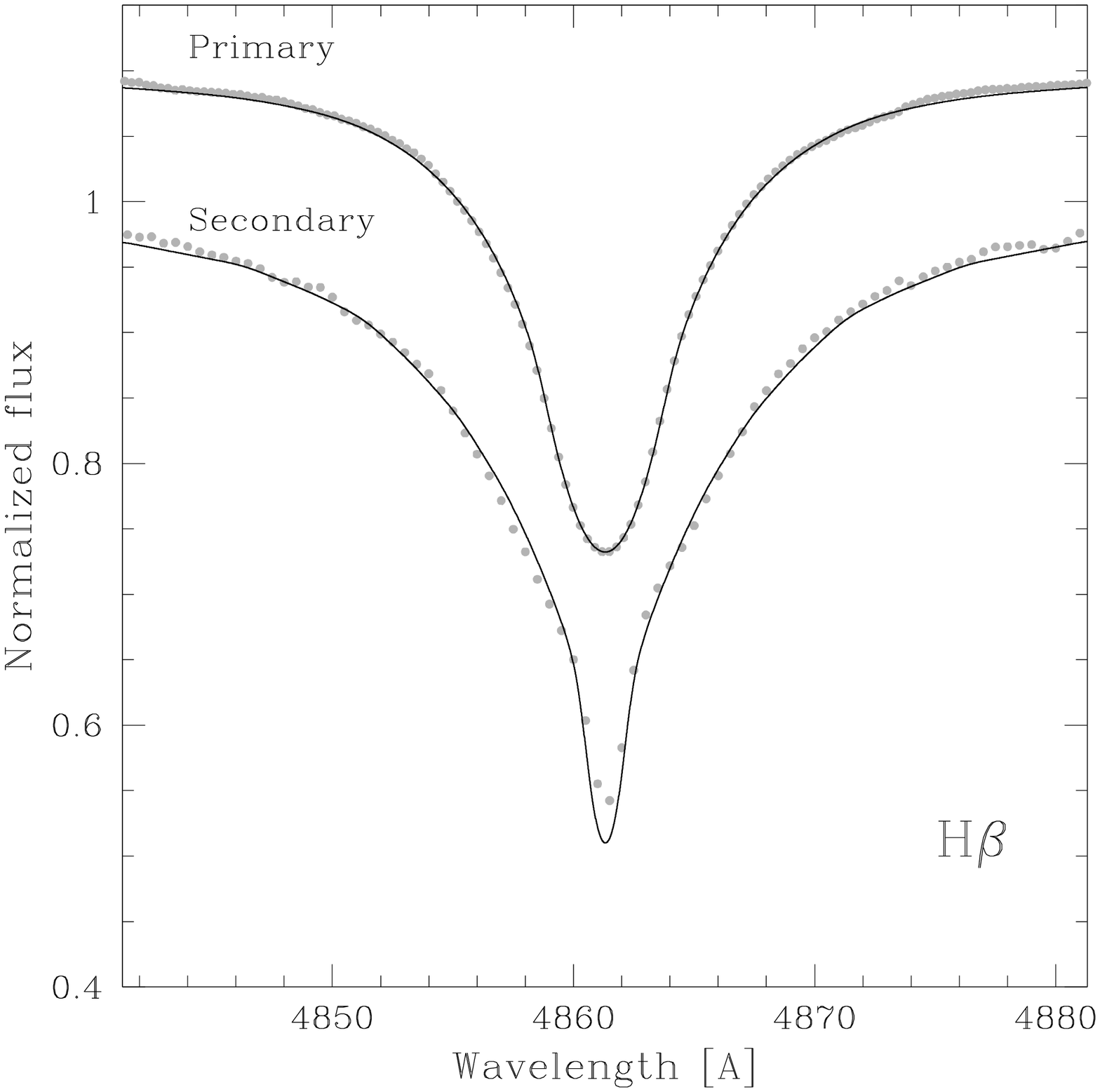}
\caption{Best fit model (solid line) to the disentangled spectra
(symbols) of both components of the Spica binary system. The
spectrum of the secondary was vertically shifted for clarity.}
\label{Figure6}
\end{figure*}

\begin{figure*}
\includegraphics[scale=0.40]{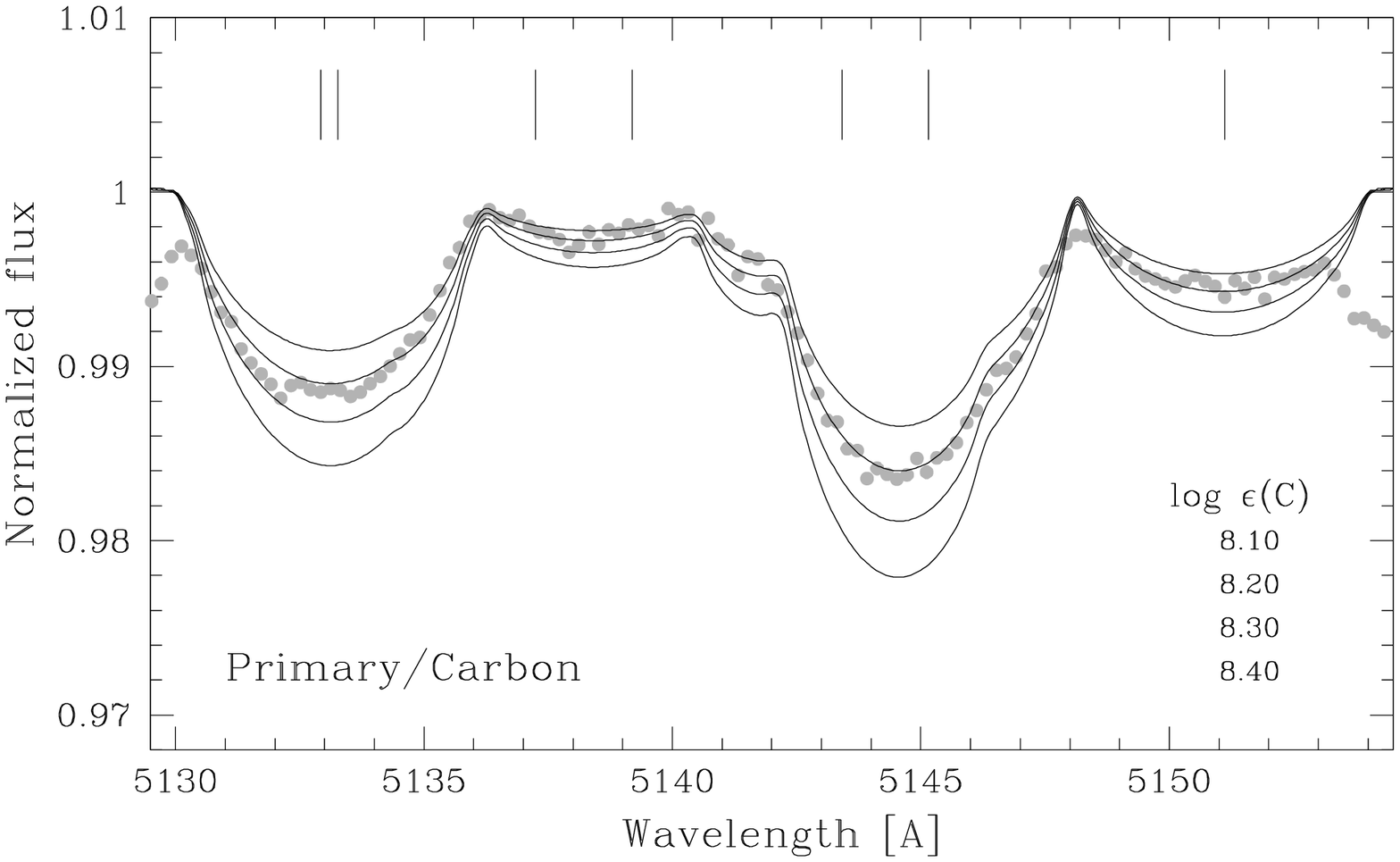}\hspace{5mm}
\includegraphics[scale=0.40]{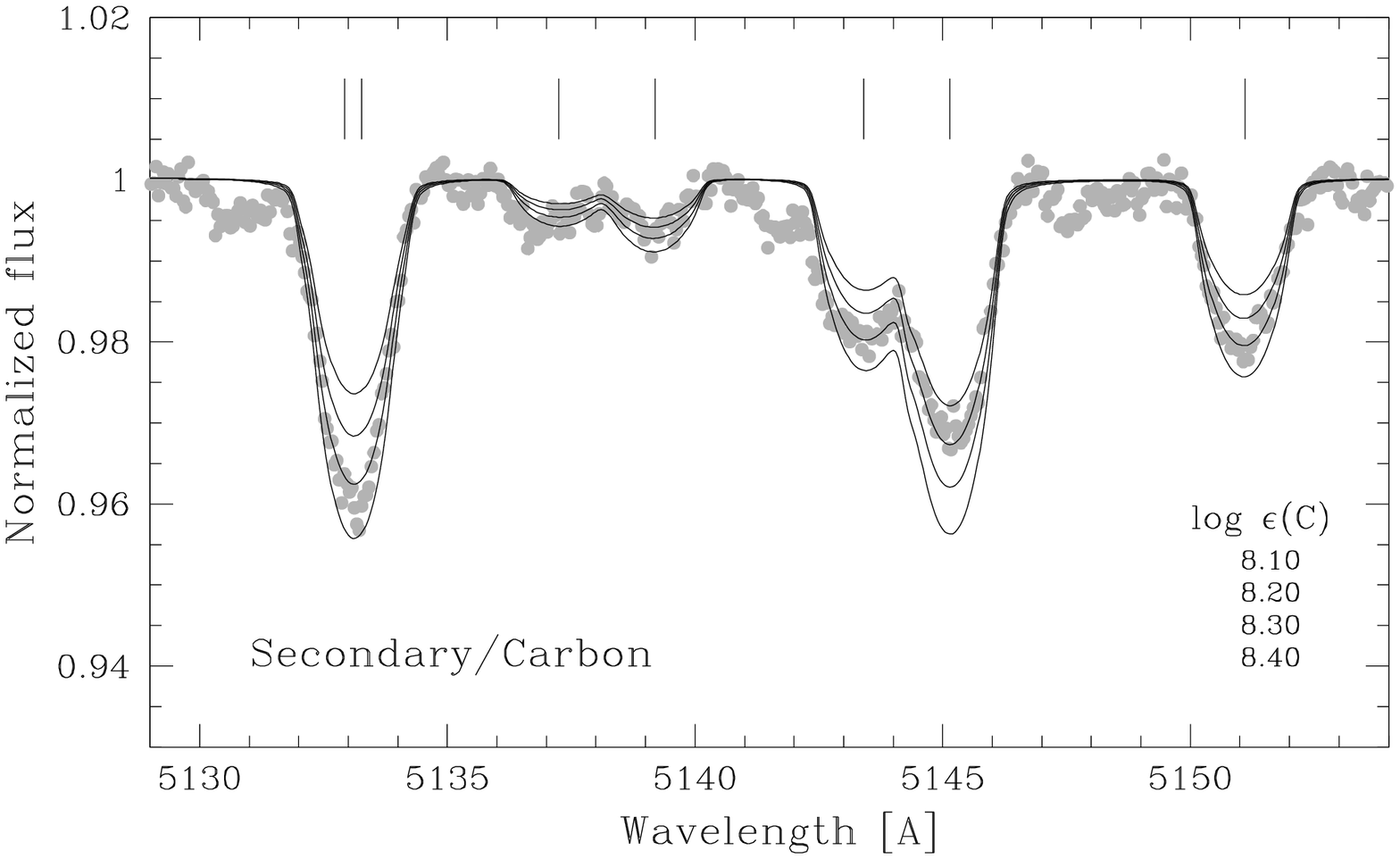}\vspace{5mm}
\includegraphics[scale=0.40]{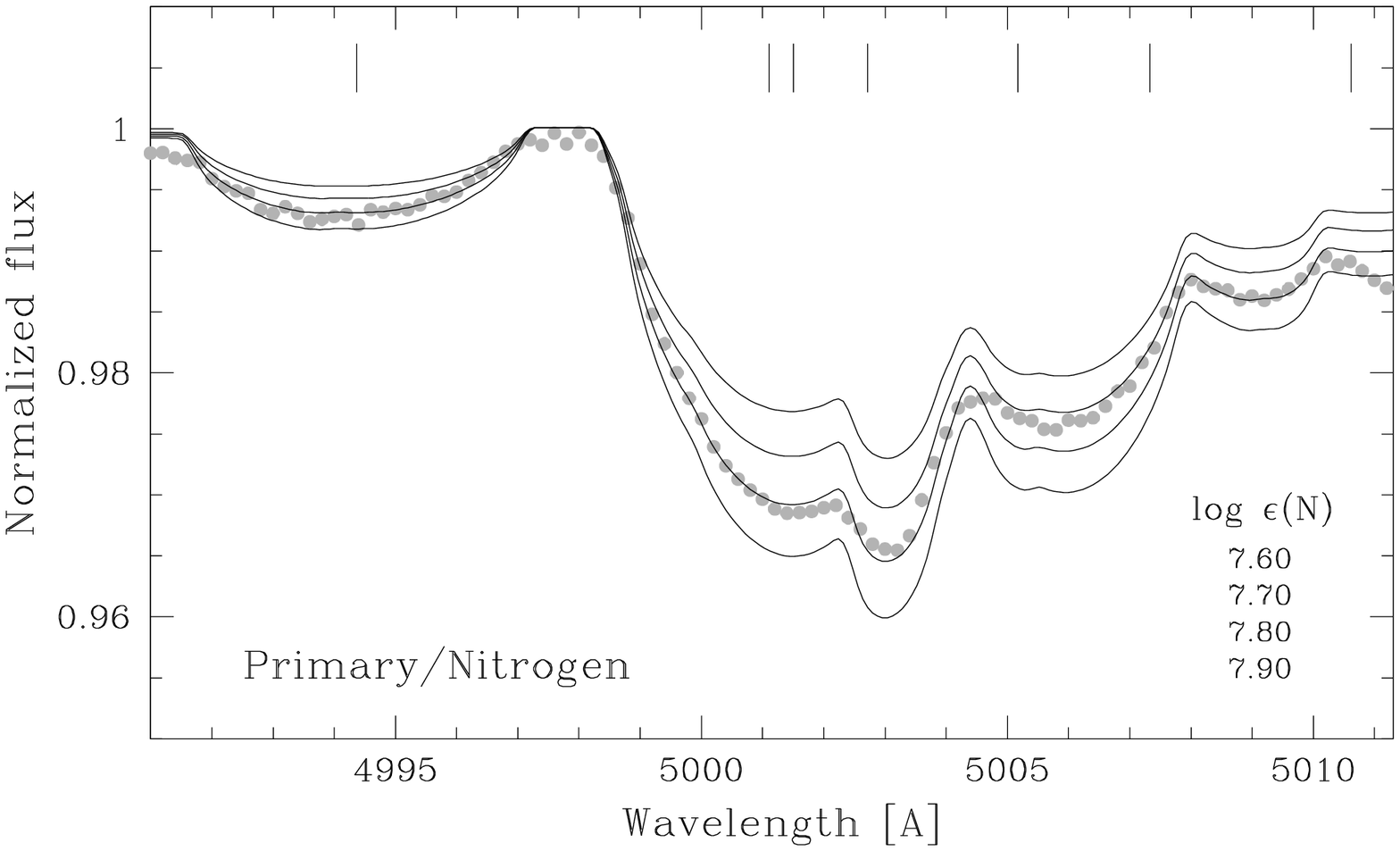}\hspace{5mm}
\includegraphics[scale=0.40]{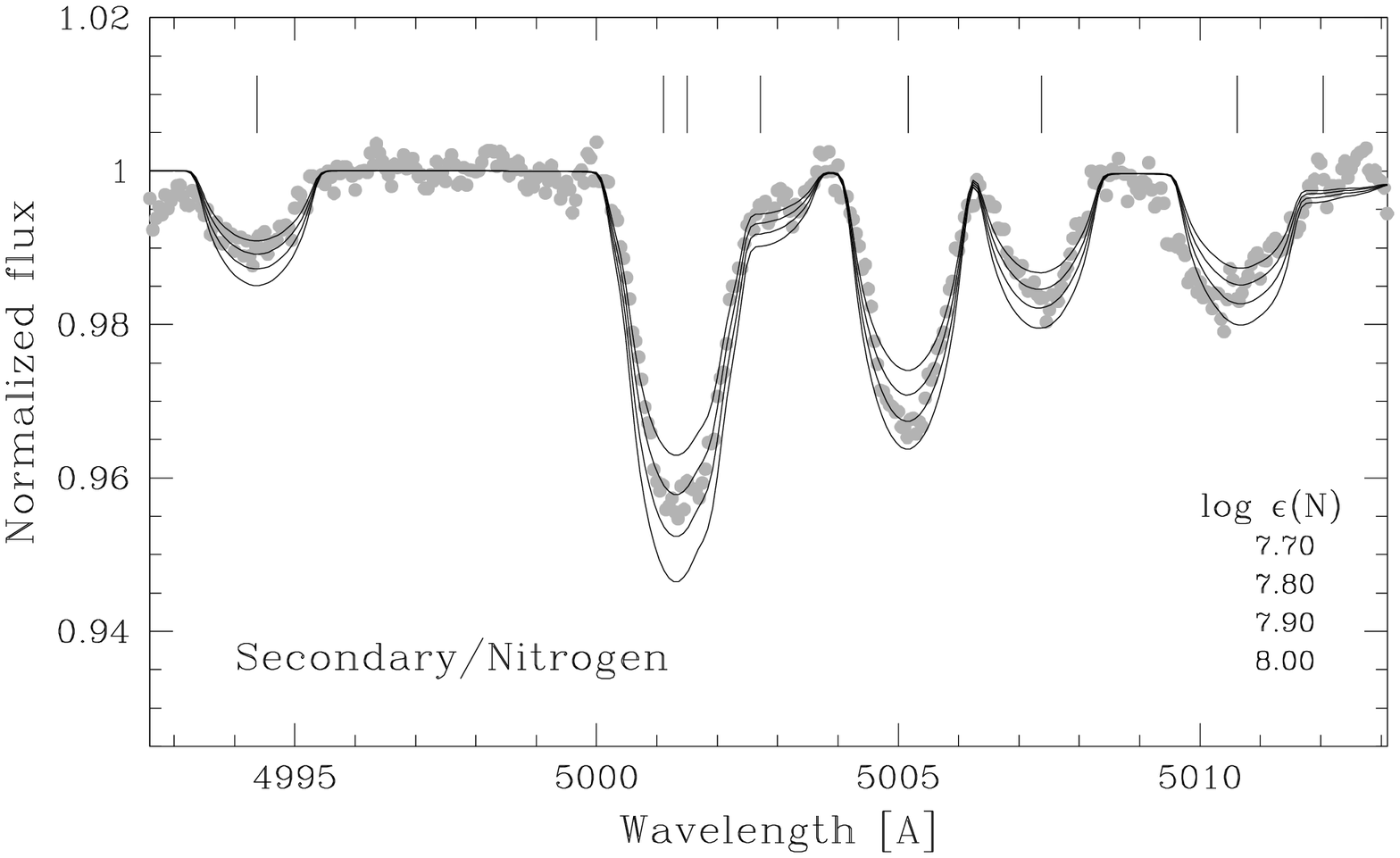}
\caption{Fits of the theoretical spectra (lines) to the renormalised
disentangled spectra of both stars (symbols) in the regions of
carbon (top) and nitrogen (bottom) lines, and assuming different
abundances of these elements. The columns (from left to right) refer
to the primary and secondary component; the assumed abundances are
indicated in the plot, where the smallest number corresponds to the
weakest line. Note different X and Y scales for the primary and
secondary which is due to large difference in \vsini\ values of the
two stars.} \label{Figure7}
\end{figure*}

Before the spectra of individual binary components can be analysed
to determine basic atmospheric parameters and individual abundances,
they need to be renormalised to the individual continua of the
component stars. This is done by means of the light ratio, which is
often determined from photometric data for eclipsing binaries and by
means of the constrained fitting \citep[i.e., fitting for the light
ratio simultaneously with other fundamental
parameters,][]{Tamajo2011} for non-eclipsing systems. In the
particular case of Spica, we fit for the light ratio and the other
atmospheric parameters of the individual stellar components
simultaneously, so that the light contribution of each of the stars
is one of the output parameters from our spectrum analysis. The
fundamental parameters and chemical composition of both stars were
derived by fitting their disentangled spectra to a grid of synthetic
spectra, based on the so-called hybrid approach. The latter assumes
the usage of LTE-based atmosphere models and non-LTE spectral
synthesis, in our case computed with the {\sc atlas9}
\citep{Kurucz1993}, and {\sc detail} \citep{Butler1984} and {\sc
surface} \citep{Giddings1981} codes, respectively. Justification of
such approach is discussed in \citet{Nieva2007}. The adopted
spectrum analysis procedure is discussed in
\citet{Tkachenko2014a,Tkachenko2014b} in full detail but here we
summarize a few key points only: 1) the whole analysis is based on
the method of spectrum synthesis where a grid of synthetic spectra
is fitted to the observations. Not only this concerns the
determination of the atmospheric parameters of the individual
stellar components but also their chemical compositions: both stars
exhibit significant line broadening in their spectra, so the method
of spectral synthesis is superior for the determination of
individual abundances compared to the equivalent width measurements
which suffers from heavy blending of spectral lines. 2) The
determination of atmospheric parameters such as \te\ and \logg, as
well as the individual light factors is largely based on fitting the
Balmer lines (H$_{\delta}$, H$_{\gamma}$, and H$_{\beta}$). 3)
Microturbulent velocities are determined by fitting selected oxygen
lines requiring a null correlation between the individual abundances
and the equivalent widths of the corresponding spectral lines. The
lines selected are those from \citet{Simon-Diaz2010}. We also use
silicon lines as an additional check of the reliability of the
determined microturbulent velocity. The projected rotational
velocities of the stars are determined by fitting the metal line
spectrum of each of the components.

Tables~\ref{Table4} and \ref{Table5} list the atmospheric parameters
and individual abundances for both stellar components of the Spica
binary system. The primary component is found to be hotter and more
evolved than its companion, whereas both stars have the same
chemical composition within the quoted error bars. The errors are
1$\sigma$ uncertainties, that for individual abundances take into
account line-to-line scatter and the error propagation from the
atmospheric parameters. The spectroscopically derived effective
temperature of the secondary and surface gravities of both stars are
in good agreement with the corresponding values of $T_{\rm
eff,2}$=20\,585$\pm$850~K, $\log{g_1}$=3.75$\pm$0.11~dex, and
$\log{g_2}$=4.15$\pm$0.16~dex obtained from the light curve fitting
and the dynamical masses and radii of the stars, respectively. The
same concerns light factors: the light curve solution delivers
contributions of 84.4/15.6\% for the primary/secondary, which is in
agreement with the spectroscopic findings within about 1\%. Our
atmospheric parameters are also in good agreement with the values
reported by \citet{Lyubimkov1995}, who found $T_{\rm
eff}$=24\,700$\pm$500/20\,800$\pm$1\,500~K,
$\log{g}$=3.7$\pm$0.1/4.2$\pm$0.2~dex, and
\vsini=161$\pm$2/70$\pm$5~\kms\ for the primary/secondary component.
Any discrepancies between our parameters and those determined by
\citet{Lyubimkov1995} can be attributed to the fact that, in the
latter study, the analysis was based on the original, composite
spectra of the binary. The approach implemented in the current study
and that is based on the analysis of the disentangled spectra of
both components is more accurate as it does not suffer from spectral
line blending due to the contribution of a companion star.

The quality of the fit to three Balmer lines (H$_{\delta}$,
H$_{\gamma}$, and H$_{\beta}$) in the spectra of both stars is shown
in Fig.~\ref{Figure6}. A set of carbon and nitrogen lines in the
spectra of both binary components is illustrated in
Fig.~\ref{Figure7}, along with the theoretical spectra computed
assuming different abundances of these elements. A significant
difference in rotational broadening of the spectral lines of the
stars is easily caught by eye, as well as the similarity in their
chemical compositions. Finally, we note that chemical compositions
of both stars found in this study agree within the error bars with
both the Cosmic Abundance Standard \citep{Nieva2012} and chemical
composition of the Sun as derived by \citet{Asplund2009}. An
exception is the abundances of carbon and aluminium in the
atmosphere of the primary component, which seem to be slightly
overabundant given our 1$\sigma$ uncertainties. We do not find any
increase in the surface nitrogen abundance for either of the binary
components, which puts both stars in the Group 1 of \citet[][Figure
1]{Hunter2008}. For the further analysis, we will assume solar
chemical composition for both binary components, as the above
mentioned deviations of carbon and aluminium found for the primary
will not have any impact on the analysis outlined in the next
sections.

\begin{table*} \tabcolsep 1.5mm\caption{\small Results of the frequency analysis. The Rayleigh limit amounts to 0.044~\cd\ for the photometry and 0.008~\cd\ for the spectroscopy. Errors in the amplitude are given in parenthesis in terms of last
digits. The photometric frequencies which are in common with the
spectroscopic ones are highlighted in boldface.}
\begin{tabular}{lllll|llll|llll} \hline
\multicolumn{5}{c|}{Photometry\rule{0pt}{9pt}} &
\multicolumn{8}{c}{Spectroscopy}\\
\hline
 & & & & & \multicolumn{4}{c|}{Pixel-by-pixel\rule{0pt}{9pt}} & \multicolumn{4}{c}{RV}\\
\multicolumn{5}{c|}{(magnitude)} & \multicolumn{4}{c|}{(continuum
units)} & \multicolumn{4}{c}{(\kms)}\\ \hline f$_i$ & Freq (\cd) &
Freq (\mhz) & Amplitude & S/N & Freq\rule{0pt}{9pt} (\cd) & Freq
(\mhz) & Amplitude & S/N & Freq (\cd) & Freq (\mhz) & Amplitude & S/N\\
\hline
f$_1$ & 0.694\rule{0pt}{9pt} & 8.030 & 0.00198(7) & 6.9 & 3.001 (12f$_{\rm orb}$) & 34.722 & 0.014(1)& 10.8 & 2.998 (12f$_{\rm orb}$) & 23.117 & 4.32(3) & 12.1\\
f$_2$ & 0.638 & 7.382 & 0.00114(7) & 5.5 & 5.748 & 66.504 & 0.012(2) & 7.4 & 5.751 & 66.539 & 1.93(3) & 4.5\\
f$_3$ & 1.486 & 17.193 & 0.00108(5) & 5.0 & 0.252 (f$_{\rm orb}$) & 2.916 & 0.011(1) & 5.6 & 0.248 & 2.869 & 1.4 & 4.1\\
f$_4$ & {\bf 0.254} (f$_{\rm orb}$) & {\bf 2.939} & 0.00083(5) & 4.4 & 2.162 & 25.014 & 0.007(1) & 4.0 & --- & --- & --- & ---\\
f$_5$ & 0.741 (3f$_{\rm orb}$) & 8.573 & 0.00082(5) & 4.1 & & & & & &\\
f$_6$ & {\bf 2.149} & {\bf 24.864} & 0.00063(4) & 4.0 & & & & & & & &\\
f$_7$ & {\bf 5.755} & {\bf 66.585} & 0.00057(3) & 5.5 & & & & & & & &\\
\hline
\end{tabular}
\label{Table6}
\end{table*}

\section{Frequency analysis and mode identification}\label{Sect: frequency analysis}

In this section, we address the question of the variability
intrinsic to either of the binary components, that manifests itself
in terms of the brightness and line profile variations. The results
of spectroscopic mode identification are also presented and
discussed.

\subsection{Previous work}

The intrinsic variability of the primary component of Spica was
extensively studied in the past. \citet{Shobbrook1969} were the
first to report on the detection of light variations intrinsic to
the evolved primary with frequency of $\sim$5.75~\cd\ (66.53~\mhz).
The variability has been attributed to a radial pulsation mode, thus
the star has been classified as a $\beta$~Cep-type variable.
\citet{Smak1970} came to the same conclusion and reported about a
steady decrease of the dominant pulsation period at a rate of about
5 seconds per century. \citet{Shobbrook1972} investigated the system
based on newly obtained photometric and spectroscopic data and
confirmed the primary to be a variable star with the dominant
frequency of $\sim$5.75~\cd\ (66.53~\mhz). However, the authors
found the pulsation mode to be variable in amplitude with the
variability being random. A second pulsation mode was detected at a
frequency of $\sim$3.97~\cd\ (45.93~\mhz) in the radial velocity
(RV) data and interpreted as the fundamental radial mode. The
dominant mode was proposed to be due to the first overtone radial
pulsation.

\citet{Walker1982} investigated new spectroscopic data obtained at
orbital phases of large RV separation of the components. The authors
detected ``bumps'' moving across line profiles of the primary, just
in line with the expectations for non-radial pulsations.

A detailed study of the variability intrinsic to the primary
component was performed by \citet{Smith1985a,Smith1985b}. About 500
high resolution, high signal-to-noise-ratio (S/N) spectra were
collected and analysed in the region of the Si~{\small III} triplet
at $\lambda\lambda$ 4553, 4568, and 4575~\AA. Clear, though
low-amplitude, variability has been detected in the RV data of all
three components of the triplet. The signal was found to be due to
four modes, f$_1\sim$3.68~\cd\ (42.58~\mhz), f$_2\sim$7.5~\cd\
(86.77~\mhz), f$_3\sim$0.49~\cd\ (5.67~\mhz), and f$_4\sim$2.99~\cd\
(34.59~\mhz); no evidence of the previously reported radial mode
with frequency of $\sim$5.75~\cd\ (66.53~\mhz) was found in the
data. From the spacings between individual ``bumps'' detected in the
silicon line profiles of the primary, the modes f$_1$ and f$_2$ were
found to have geometries corresponding to $l=m=8$ and $l=m=16$,
respectively. The author also claims that these two modes have the
same, high radial order n. The third mode at 0.49~\cd\ (5.67~\mhz)
was attributed to the spectroscopic analog of the photometric
ellipsoidal variability and identified as an $l=2$ mode. It is worth
noting that this variability was detected indirectly, as a
perturbing agent on the $l=8$ mode. \citet{Smith1985a} concluded
that the system has achieved angular momentum equilibrium by
eventually rotating bisynchronously, i.e. with the rotation period
of the primary being half of the orbital period of the binary.

The f$_4\sim$2.99~\cd\ (34.59~\mhz) mode was analysed by
\citet{Smith1985b} in detail based on the data set obtained by
\citet{Smith1985a}. The author found that the variability in
question has very distinct behavior: unlike non-radial pulsations
that show continuously travelling bumps from blue to red wing of the
line profile or vice verse, this variability is characterized by
nearly stationary absorption like features in the wings. The
amplitude of these features was found to be variable, making them to
disappear and reappear again during the variability cycle.
\citet{Smith1985b} found that the variability due to the
f$_4\sim$2.99~\cd\ (34.59~\mhz) mode can be well reproduced with a
spherical harmonic function with the radial component of the
displacement vector suppressed. Thus, the mode was found to have
characteristics similar to those of toroidal modes, and was
designated by the author as a ``quasi-toroidal mode''.

\begin{figure}
\includegraphics[scale=0.72]{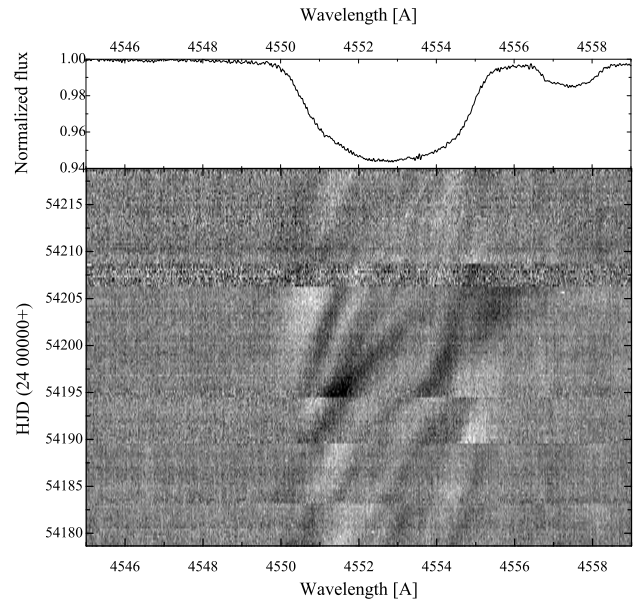}
\caption{{\bf Top:} Average profile of the Si~{\small III}
$\lambda\lambda$~4552.6~\AA\ line computed from the time series of
162 composite spectra taken within narrow range in orbital phase.
{\bf Bottom:} Time-series of the residual spectra obtained after
subtraction of the average profile. The grey scale represent the
residual intensity at each wavelength pixel.} \label{Figure8}
\end{figure}

\citet{Harrington2009} presented an analysis of high resolution
spectra taken with the {\sc espadons} instrument attached to the
Canada France Hawaii Telescope. The authors reported the detection
of the lines of the secondary component in their spectra, and
classified both components as B-type stars. The authors found all
weak spectral lines in their spectra to display discrete narrow
features which could not be identified in He~{\small I} and Balmer
lines. The authors concluded that tidal flows exerted by the
main-sequence secondary on the evolved primary are the main
contributor to the short-term variability observed in the line
profiles of the Spica system. \citet{Palate2013} arrived to a
similar conclusion by analysing the data set obtained by
\citet{Harrington2009} and using improved methodology. The
conclusions of these two papers will be discussed in the context of
our own results in the last section of the current study.

\subsection{MOST photometry}

The photometric solution obtained in Sect.~\ref{Sect: phoebe} was
used to compute residuals from the original MOST light curve
illustrated in Fig.~\ref{Figure1} (top panel). For the extraction of
individual frequencies, amplitudes, and phases from the residual
signal, we used the Lomb-Scargle version of the discrete Fourier
transform \citep{Lomb1976,Scargle1982} and consecutive prewhitening.
A detailed mathematical description of the procedure can be found in
\citet{Degroote2009}.

\begin{figure*}
\includegraphics[scale=0.72]{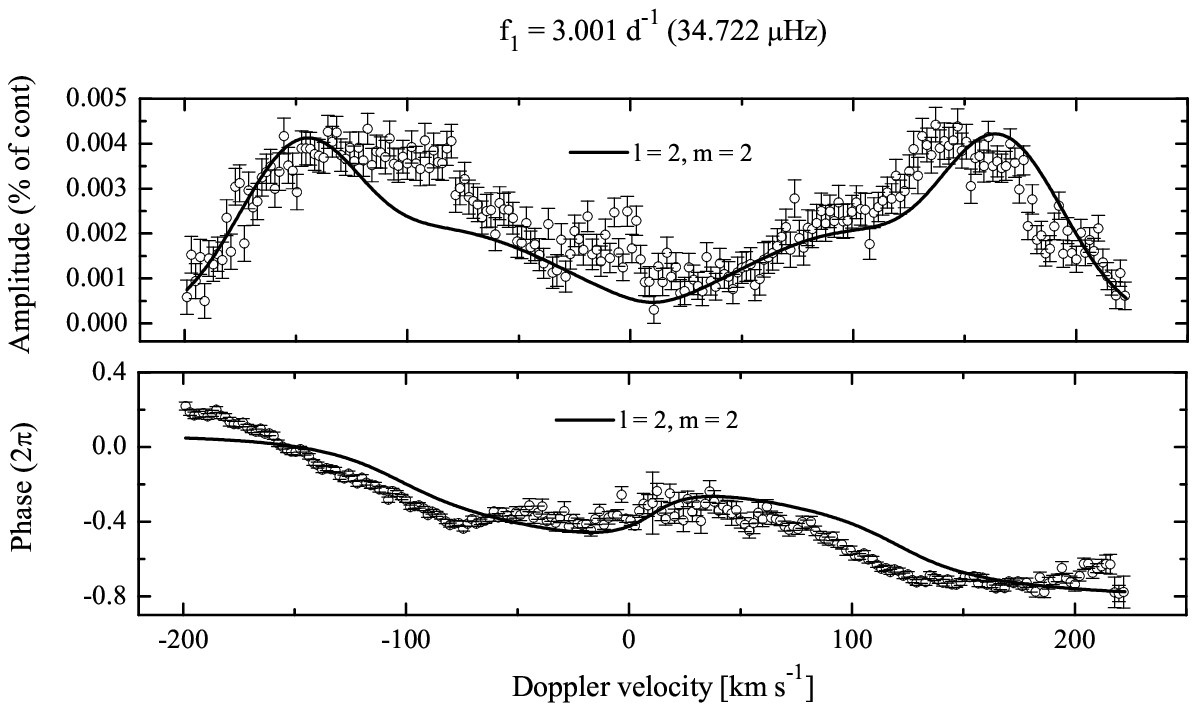}
\includegraphics[scale=0.72]{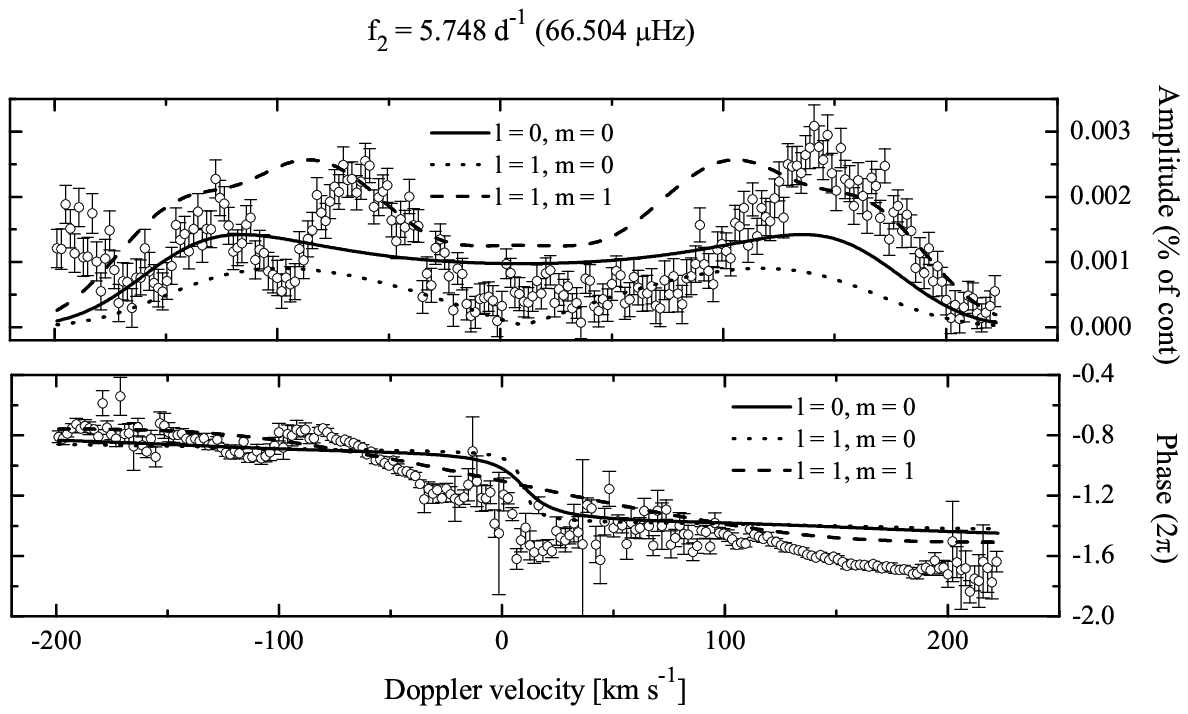}\vspace{3mm}
\includegraphics[scale=0.72]{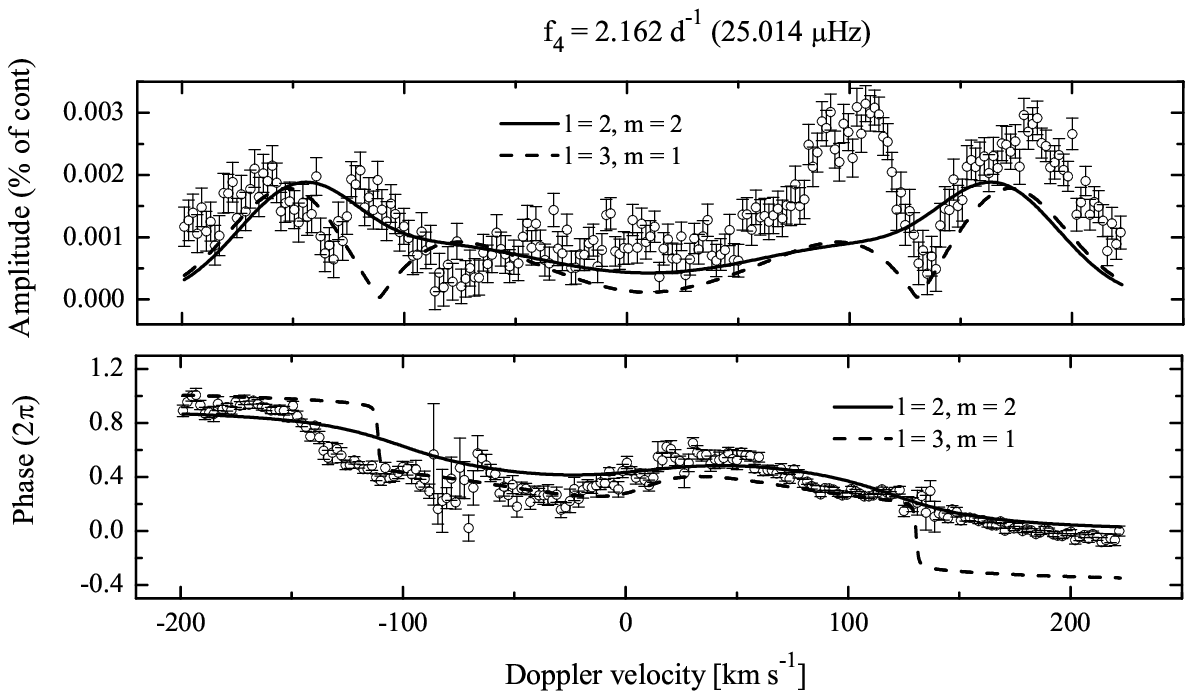}
\caption{Amplitude and phase distributions across the line profile
for three pulsations modes: f$_1=$3.001~\cd\ (top left),
f$_2=$5.748~\cd\ (top right), and f$_4=$2.162~\cd\ (bottom). The
best fit models are overplotted with the lines of different style;
the corresponding mode geometry is indicated in each subplot.}
\label{Figure9}
\end{figure*}

The list of photometric frequencies extracted from the MOST light
curve is given in Table~\ref{Table6} (first column, labeled
``Photometry''). The frequencies with values of S/N above 4 are
listed only. Those which are in common with the spectroscopic
frequencies are highlighted in boldface. Frequencies f$_4$ and f$_5$
are linked to the orbital frequency of the binary system and are due
to imperfect removal of the orbital signal from the original light
curve. Frequency f$_7$ has been previously reported in the
literature as a radial pulsation mode with (randomly) variable
amplitude \citep[e.g.,][]{Shobbrook1969,Shobbrook1972}.
\citet{Smith1985a} reported that this frequency could not be
detected in his spectroscopic data, but stressed that there is no
reason to assume that the mode will not return eventually. All other
frequencies detected by us in the MOST photometry are new and have
not been reported in the literature before. Based on the results of
our spectroscopic investigations (see below), we conclude that it is
the primary component that is responsible for the photometric
variability occurring on top of the orbital signal.

\subsection{High-resolution spectroscopy}

Our very first exercise with the spectra was to figure out which of
the stars is intrinsically variable in the system. To do so, we have
selected all individual spectra in a narrow range of orbital phase
to make sure that all dynamics we see in the line profiles is due to
variability intrinsic to either of the binary components and not due
to the orbital motion within the system. Also, if all variability
was due to tides and/or tidally locked \citep[see,
e.g.,][]{Harrington2009,Palate2013}, one would expect the signal to
be repetitive with orbital cycle, i.e. nearly the same line profile
pattern would be observed at a given orbital phase every orbital
cycle \citep[see][Figure~6]{Harrington2009}. We did the above
mentioned exercise for both quadratures, where the spectral features
of the two stars are well separated in wavelength. Since the
obtained results are essentially the same for both orbital phases,
we illustrate only one of them in Fig.~\ref{Figure8}. The top panel
shows the average Si~{\small III} $\lambda\lambda$~4552.6~\AA\ line,
where the contribution of the secondary can be clearly distinguished
from the one of the primary. The bottom panel illustrates a
time-series of the residual profiles obtained by subtracting the
average profile from the individual spectra. There are 162 spectra
shown in total; all of them were used to compute the average profile
shown in the top panel. One can clearly see bumps moving from blue
to red wing of the profile of the primary component, the signal
typical for non-radial pulsations. There is no variability in our
data that could be attributed to the secondary component.

In the next step, we used the disentangled spectrum of the secondary
obtained in Sect.~\ref{Sect: SPD} to subtract the contribution of
this star from the composite spectra of the binary. To test how good
the removal of the secondary's contribution is, we did the frequency
analysis of two different sets of the residual spectra: i) all 1731
spectra; ii) excluding the spectra where the lines of the secondary
component merge with the lines of the primary. The goal of this
exercise was to check which spurious frequencies will appear in the
data due to the imperfect subtraction of the contribution of a
companion star from the composite line profiles. We found that the
frequencies at two and three times the orbital frequency (the second
and the third harmonics of f$_{\rm orb}$) show up in the first data
set while they could not be detected in our second, orbital phase
restricted data set. Thus, we conclude that these two frequencies
are not real, and base our further analysis on the second set of 716
spectra. Interestingly, \citet{Smith1985a} reported on the detection
of 2f$_{\rm orb}$ periodicity in his spectra, and attributed it to
an $l=m=+2$ ``equilibrium'' tidal mode, which is basically a
spectroscopic equivalent of the photometric ellipsoidal variability.
This type of signal was also theoretically predicted by
\citet{Harrington2009} to occur in the Spica system due to the tidal
flows exerted by the components on each other. Our test with the
real data shows that the signal at low order harmonics of the
orbital frequency is easily introduced into the data artificially,
when the contribution of the secondary component is taken into
account inappropriately, one way or the other.

The individual frequencies have been extracted from both RVs
(computed as the first order moment of the spectral line) and the
line profiles themselves, using discrete Fourier-transform (DFT) and
a consecutive prewhitening procedure as implemented in the {\sc
famias} \citep{Zima2008} software package. Following the procedure
described in \citet{Tkachenko2014a}, the DFT was computed up to the
Nyquist frequency, the amplitudes and phases were optimised at each
step of the prewhitening procedure while keeping the frequency
values fixed, and the frequencies were accepted as significant ones
if their S/N was higher or equal to 4 \citep{Breger1993}. For the
analysis, we used the Si~{\small III} triplet $\lambda\lambda$ 4553,
4568, and 4575~\AA, but neglected all other lines potentially
sensitive to non-radial pulsations (like He or Mg lines), because of
high degree of their blending due to the rapid rotation of the star.
The final list of frequencies is given in the second column of
Table~\ref{Table6}, labeled ``Spectroscopy''. Frequencies
f$_2\sim$5.75~\cd\ (66.53~\mhz) and f$_3\sim$0.25~\cd\ (2.89~\mhz)
were found both in the RVs and in the line profiles themselves, and
agree within the error bars with the frequencies f$_7$ and f$_4$
detected in the photometric data. As it has been discussed in the
previous section already, the former is likely a radial mode whereas
the latter is the orbital frequency. There is a frequency,
f$_1\sim$3.00~\cd\ (34.71~\mhz), that shows up in both spectroscopic
observables, but could not be detected in the MOST photometry. This
one is close to the 12th harmonic of the orbital frequency and is
thus a good candidate for a mode excited by means of the dynamical
tides expected to occur in eccentric binary systems. The same
frequency was detected by \citet{Smith1985a} in his spectroscopic
data and explained in terms of a tidally forced ``quasi-toroidal''
mode \citep{Smith1985b}. The last frequency, f$_4\sim$2.16~\cd\
(24.99~\mhz), could be detected in the line profiles but not in the
RV data, and has a low amplitude both in the spectroscopic and the
photometric data.

We used {\sc famias} to identify all three modes, f$_1$, f$_2$, and
f$_4$, detected in our spectroscopic data. For that, we used the
Fourier-parameter fit method \citep[FPF,][]{Zima2006}, which is best
suitable for mode identification in rapidly rotating stars.
Identification of pulsation modes requires a knowledge of
fundamental stellar parameters like mass, radius, effective
temperature, surface gravity, etc. These were set to the values
obtained from our combined photometric and spectroscopic solution,
as summarized in the third column of Table~\ref{Table7}. The
individual masses and radii of the stars were computed from binary
dynamics and we refer the reader to \citet[][Section
5]{Tkachenko2014b} for any details. The luminosities were computed
from the dynamical radii and effective temperatures of stars. In the
first step, the modes were identified independently of each other,
that is each time the star was treated as a mono-periodic pulsator.
This often allows to narrow the range in free parameters,
particularly in $l$ and $m$ quantum numbers, which in turn
significantly reduces the calculation time when the multi-periodic
solution is searched for. To provide sufficient spatial resolution,
we divided the stellar surface into 10 000 segments. At each
iteration step the amplitude and phase distributions across the
profile were optimized along with the zero-point profile. Besides
the $l$ and $m$ quantum numbers and intrinsic amplitudes of the
modes, the intrinsic width of the Gaussian profile $\sigma$,
projected rotational velocity of the star \vsini, and the
inclination of its rotational axis $i_{\rm rot}$ were set as free
parameters. Since {\sc famias} assumes an alignment between rotation
and pulsation axes, it is possible to check for possible spin-orbit
misalignment by keeping $i_{\rm rot}$ as a free parameter.
Figure~\ref{Figure9} illustrates the distributions of amplitude and
phase across the line profile for all three modes: f$_1=$3.001~\cd\
(top left), f$_2=$5.748~\cd\ (top right), and f$_4=$2.162~\cd\
(bottom). The best fit models with $\chi^2$ values below 20 are
overplotted with the lines of different style; the corresponding
mode geometries are indicated in the plots. Such large values of
$\chi^2$, that are obviously linked to the inconsistency between the
model and observations in Fig.~\ref{Figure9}, can be attributed to
the fact the high-order rotation effects are not taken into account
in the {\sc famias} model. The f$_1=$3.001~\cd\ (34.722~\mhz) mode
is unambiguously identified as an $l=m=2$ mode. No unique solution
could be found for the modes f$_2$ and f$_4$, though the range in
both quantum numbers is well constrained: we find that the former
mode is likely a radial or $l=1$ mode, while the latter is either
$l=2$ or $l=3$ mode. All our best solutions with $\chi^2$ values
below 20 suggest a range in the inclination angle of rotation axis
$i_{\rm rot}$ between 53 and 66 degrees. This value is in good
agreement with the orbital inclination of $i_{\rm orb}=63.1$
degrees.

\begin{table}
\tabcolsep 2.0mm\caption{Fundamental stellar parameters of both
components of the Spica binary system derived from the combined
photometric and spectroscopic solution.}\label{Table7}
\begin{tabular}{llr@{$\pm$}lr@{$\pm$}l} \hline
Parameter & Unit & \multicolumn{2}{c}{Primary} &
\multicolumn{2}{c}{Secondary}\\
\hline
Mass, $M$\rule{0pt}{9pt} & M$_\odot$ & 11.43&1.15 & 7.21&0.75\\
Radius, $R$ & R$_\odot$ & 7.47&0.54 & 3.74&0.53\\
Luminosity, $\log L$ & L$_\odot$ & 4.312&0.095 & 3.353&0.181\\
Effective temperature$^1$, \te & K & 25\,300&500 & 20\,900&800\\
Surface gravity$^1$, \logg & dex & 3.71&0.10 & 4.15&0.15\\
\hline\multicolumn{3}{l}{$^1$ adopted from spectroscopy (see
Table~\ref{Table4})}
\end{tabular}
\end{table}



\section{Evolutionary models}\label{Sect: evolutionary models}

The {\sc mesa} stellar structure and evolution code
\citep{Paxton2011,Paxton2013} was used to compute evolutionary
models for both components of the Spica system. Given that the
secondary component rotates at about 10\% of its critical velocity,
we have chosen to compute non-rotating models for this star.
Contrary, the primary is found to rotate at $\sim$30\% of its
break-up velocity, and thus the rotation could not be neglected in
the evolutionary model calculations for this binary component. The
initial abundance fractions (X, Y, Z) = (0.710, 0.276, 0.014) are
those from \citet{Nieva2012}, in agreement with the spectroscopic
findings (cf. Sect.~5). Convective core overshoot is described with
an exponentially decaying prescription of \cite{Herwig}. The Ledoux
criterion is used in the convection treatment. The OPAL opacity
tables \cite{Opal} and MESA equation-of-state are used.

\begin{figure}
\includegraphics[scale=0.57]{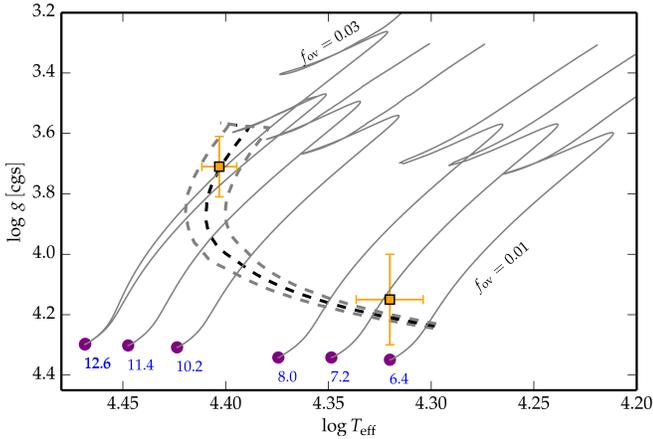}
\caption{Location of the primary and secondary components of the
Spica system in the \te-\logg\ Kiel diagram, along with the {\sc
mesa} evolutionary tracks. A track corresponding to a higher
overshoot is shown for one selected mass to illustrate that it is
impossible to constrain overshoot parameter $f_{ov}$ from the models
due to large error box in the determined fundamental parameters.
Dashed lines show three isochrones corresponding (from bottom to
top) to 11.5, 12.5, and 13.5~Myr} \label{Figure10}
\end{figure}

Figure~\ref{Figure10} illustrates the position of both stellar
components of the Spica system in the \te-\logg\ diagram. The error
bars are those obtained from 1$\sigma$ combined spectroscopic and
photometric uncertainties (cf. Table~\ref{Table4}). The evolutionary
tracks are shown with solid lines for dynamical masses of both stars
and the corresponding error bars (cf. Table~\ref{Table7}). The
tracks were computed for the overshoot parameter $f_{ov}=0.01$
($\alpha_{ov}\approx0.1$~H$_p$); we also show the track computed
with $f_{ov}=0.03$ ($\alpha_{ov}\approx0.3$~H$_p$) for a single mass
value to illustrate the effect of overshoot on the evolutionary path
of the star. Dashed lines show isochrones corresponding to 11.5
(bottom), 12.5 (middle), and 13.5 (top) Myr. The latter were
computed assuming overshoot of $f_{ov}=0.01$. We find that the
position of the unevolved secondary agrees very well with its
dynamical mass evolutionary track. The same holds for the primary
when the error bars on the mass, \te, and \logg\ of the star are
taken into account. The positions of both stars are in good
agreement with the isochrone corresponding to 12.5~Myr. By taking
into account the errors on \te\ and \logg\ for the primary
component, we estimate the accuracy for the age to be of $\pm1$~Myr.

\section{Summary and Discussion}

In this paper, we presented a detailed analysis of the Spica close
binary system, based on high-resolution spectroscopy obtained with
the {\sc coralie} spectrograph and space-based photometry gathered
with the MOST mission. In total, 1731 spectra and $\sim$23~days of
nearly continuous photometric measurements have been analysed with
the state-of-the-art modelling techniques. We found that the Spica
system is close to showing a grazing secondary eclipse, the fact
exciting by itself but hardly helpful for the analysis in terms of a
better precision on the fundamental parameters of the component
stars. Both binary components are found to have similar chemical
composition consistent with the cosmic abundances derived from
Galactic OB stars \citep{Nieva2012}.

A comparison of the positions of both stars in the \te-\logg\
diagram with the {\sc mesa} evolutionary models reveals a good
agreement for the secondary component. This result differs from the
one obtained by us for the V380\,Cyg \citep{Tkachenko2014a} and
Sigma Scorpii \citep{Tkachenko2014b} systems, where the unevolved
secondary components were found to show significant discrepancy
between their evolutionary and dynamical masses. A small mass
discrepancy is not excluded for the primary component of Spica,
though no definite conclusion can be drawn on this aspect because of
rather large uncertainty of about 10\% in the stellar mass.

Contrary to \citet{Harrington2009} and \citet{Palate2013} suggesting
that the line profile variations observed in the primary component
can be naturally explained in terms of surface tidal flows exerted
on it by the secondary, we find a clear evidence of non-radial
pulsation modes in the system. One of the modes is interpreted by us
as being excited through a resonance between the free oscillation of
the primary and the dynamical tides in the binary system. If
confirmed with future observations, the Spica system will be the
first massive binary (with the mass of the pulsating component
larger than 5~M$_{\odot}$) in which tidally-induced pulsations have
been detected. Unfortunately, our attempt to carry out detailed
asteroseismic modelling for the primary component of the Spica
system was unsuccessful due to insufficient frequency resolution
achieved with our data. In the result, several tens of models with
$\chi^2$ values below 1 were found when attempting to fit the
observed frequencies, implying that the actual errors on individual
frequencies are larger than the difference between the observed and
theoretical values. The corresponding models cover entire parameter
range in the fundamental stellar parameters such as the effective
temperature, surface gravity, mass, radius, etc., implying that
detailed asteroseismic analysis is not feasible with the data set we
currently have. An accumulation of additional spectroscopic material
would help to increase the frequency resolution, and thus would make
the seismic analysis feasible for the primary component of Spica.

The mass discrepancy problem remains an unresolved issue in the
massive binary research area for at least a few decades. It clearly
points to some shortcomings in the current theories of stellar
structure and evolution since the masses of the component stars
measured from binary dynamics are model-independent values. As of
today, the mass discrepancy problem, one way or the other, was
addressed in several studies
\citep[e.g.,][]{Burkholder1997,Pavlovski2009,Tkachenko2014a,Tkachenko2014b,Garcia2014}.
The majority of the studies reported on a significant discrepancy
between the dynamical and evolutionary masses, in a very few cases,
the models were successful in matching the dynamical masses of the
component stars. Revealing the shortcomings in the current theories
of stellar structure and evolution requires a statistically
significant, homogeneously analysed sample of massive binaries,
whereas currently the analysis methods as well as the evolutionary
models used by different authors are quite diverse. Spica is the
third massive binary system, after V380\,Cyg and $\sigma$\,Sco, that
has been analysed by us in a consistent, homogeneous way, using the
same modelling tools as for the other two binaries. We will extend
our sample of massive binaries by adding about a dozen of eclipsing
systems that we have recently discovered in the Campaign 0 data of
the K2 mission. More systems are expected to be found in other
fields of view of the K2 missions as well. With a few dozens of the
systems in total analysed in a consistent way, we will be able to
get more insight into the nature of the mass discrepancy problem for
massive stars.

\section*{acknowledgements}
The research leading to these results has received funding from the
European Community's Seventh Framework Programme FP7-SPACE-2011-1,
project number 312844 (SPACEINN), and from the Fund for Scientific
Research of Flanders (FWO), Belgium, under grant agreement
G.0B69.13. E.M. has received funding from the People Programme
(Marie Curie Actions) of the European Union's Seventh Framework
Programme FP7/2007-2013/ under REA grant agreement n$^\circ$623303
for the project ASAMBA. K.P. was (partially) supported by the
Croatian Science Foundation under the grant 2014-09-8656. K.Z.
acknowledges support by the Austrian Fonds zur Foerderung der
wissenschaftlichen Forschung (FWF, project V431-NBL). Mode
identification results with the software package {\sc famias}
developed in the framework of the FP6 European Coordination Action
HELAS (http://www.helas-eu.org/). A.T. dedicates this work to his
grandfather, A. Solomchenko, who passed away in January 2016.

{}

\label{lastpage}

\end{document}